\documentclass[sigconf]{acmart}

\usepackage{amsmath,amsfonts}
 
\usepackage{amssymb}

\usepackage{wrapfig} 
\usepackage{bbding} 

\usepackage{lipsum}
\usepackage{tikz}
\usetikzlibrary{arrows.meta}
\usetikzlibrary{automata,positioning}

\usepackage{threeparttable}

\definecolor{tabred}{RGB}{230,36,0}%
\definecolor{tabgreen}{RGB}{0,116,21}%
\definecolor{taborange}{RGB}{250,124,30}%
\definecolor{tabbrown}{RGB}{171,70,0}%
\definecolor{tabyellow}{RGB}{251,253,169}%
\newcommand*{\vcorr}{%
  \vadjust{\vspace{-\dp\csname @arstrutbox\endcsname}}%
  \global\let\vcorr\relax
}%

\usepackage{mathtools}

\usepackage{lscape}
\usepackage[figuresright]{rotating}
\usepackage[graphicx]{realboxes}
\usepackage{adjustbox}
\usepackage{caption}

\usepackage{subfigure}

\usepackage{colortbl} 
\usepackage{xfrac}

\usepackage{appendix}
\usepackage{float}

\usepackage{fbox} 
\usepackage{fancybox}
\usepackage{framed}

\usepackage{multirow}

\def\BibTeX{{\rm B\kern-.05em{\sc i\kern-.025em b}\kern-.08em
    T\kern-.1667em\lower.7ex\hbox{E}\kern-.125emX}}
    
\usepackage{CJKutf8}  
\usepackage{pgfplots}
\usepackage{filecontents}

\usepackage{hyperref}

\usepackage{tabularx,makecell, caption}
\usepackage{enumitem} 

\newcolumntype{L}{>{\arraybackslash}X}

\usepackage{multicol}
\usepackage{listings}
\usepackage{blindtext}
\usepackage{etoolbox,xstring,mfirstuc,textcase}
\usepackage{listings}

\usepackage{subfiles}
\usepackage[most]{tcolorbox}

\usepackage{xcolor}
\theoremstyle{plain}                
\theoremstyle{definition}       

\usepackage{siunitx}
\usepackage{comment}
\usepackage{indentfirst} 
\usepackage{framed} 

\usepackage[font=small,labelfont=bf,tableposition=top]{caption}
\usepackage{booktabs}
\usepackage{dashbox}

\usepackage{listings}
\usepackage{url}

\usepackage{color}
\usepackage{algorithm,algpseudocode} 

\renewcommand\footnotetextcopyrightpermission[1]{} 
\begin{document}
\title{Exploring Unfairness on Proof of Authority: Order Manipulation Attacks and Remedies}



\author{Qin Wang$^{2,4}$$^\star$, Rujia Li$^{1,3}$$^\star$, Qi Wang$^{1}$, Shiping Chen$^{4}$, Yang Xiang$^{2}$}\thanks{$^\star$ These authors contributed equally to the work.\\ $^\eth$ This paper has been accepted by AsiaCCS 2022.}

\affiliation{
\textit{$^1$Southern University of Science and Technology, China}\\
\textit{$^2$Swinburne University of Technology, Australia} \\
\textit{$^3$University of Birmingham, UK} \\
\textit{$^4$CSIRO Data61, Australia} 
}

\begin{abstract}
Proof of Authority (PoA) is a type of permissioned consensus algorithm with a fixed committee. PoA has been widely adopted by communities and industries due to its better performance and faster finality. In this paper, we explore the \textit{unfairness} issue existing in the current PoA implementations. We have investigated 2,500+ \textit{in the wild} projects and selected 10+ as our main focus (covering Ethereum, Binance smart chain, etc.). We have identified two types of order manipulation attacks to separately break the transaction-level (a.k.a. transaction ordering) and the block-level (sealer position ordering) fairness. Both of them merely rely on honest-but-\textit{profitable} sealer assumption without modifying original settings. We launch these attacks on the forked branches under an isolated environment and carefully evaluate the attacking scope towards different implementations. To date (as of Nov 2021), the potentially affected PoA market cap can reach up to $681,087$ million USD. Besides, we further dive into the source code of selected projects, and accordingly, propose our recommendation for the fix. To the best of knowledge, this work provides the first exploration of the \textit{unfairness} issue in PoA algorithms.

\end{abstract}


\maketitle

\section{Introduction}

\begin{figure}[!]
    \centering
    \includegraphics[width=\linewidth]{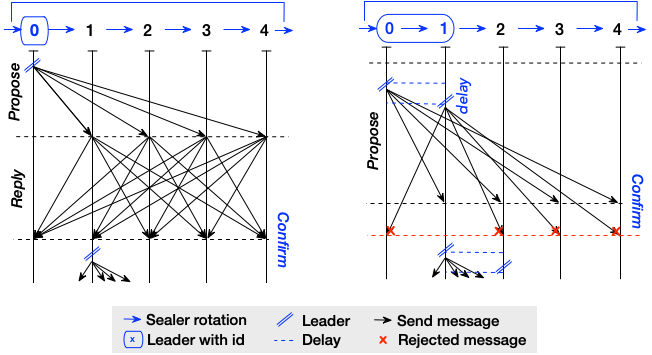}
   \caption{PoA Algorithms: \textit{Aura} (L) and \textit{Clique} (R)}
    \label{fig-poa-algm}
\end{figure}

Proof of Authority (PoA)~\cite{poachains,de2018pbft} is a type of permissioned consensus algorithm that provides a practical and effective solution for current blockchain systems, especially for consortium blockchains. The algorithm was first proposed by Wood and deployed by Ethereum \cite{wood2014ethereum} in 2017. PoA is based on a limited number of authority nodes, called \textit{sealer}s, to perform consensus procedures. Blocks and transactions have to be verified by these eligible nodes. The design of a small size committee enables fast confirmation of transactions and easy management of involved members. Many blockchain projects thereby adopt PoA as their consensus algorithms. The most famous application is Ethereum, implemented in the form of two different types~\cite{de2018pbft}: \textit{Aura} (short for Authority Round) in Parity~\cite{parity}, and \textit{Clique} in Geth \cite{geth}. \textit{Aura} modifies the traditional Byzantine Fault Tolerance (BFT)-style protocols by simplifying them into a two-phase protocol (which correspondingly requires stronger assumptions). \textit{Aura} is further adopted by Kovan Testnet \cite{kovan}, Laava \cite{weber2019platform}, VeChainThor \cite{vechain}, Microsoft Azure (deployment) \cite{azure}, and xDai DPOS network \cite{barinov2019proof}. \textit{Clique} \cite{clique} extends the permissioned settings into permissionless, by relying on the proof-of-work (PoW) consensus. It adds the procedure of authentications and brings the role of leaders. This version has been (or attempted to be) applied to Binance smart chain \cite{BSC}, Amazon blockchain network \cite{aws-Kaleido}, POA network \cite{poa}, Rinkeby \cite{rinkeby}, Quorum \cite{quorum}, Görli network \cite{goerli2022net}, Apla blockchain \cite{Apla} and HPB \cite{hpb-poa}. However, despite its wide adoption, PoA and its variants have not seriously proved to be secure. This motivates our investigation, starting with the following observations:

\smallskip
\textbf{We firstly dive into the mechanisms of PoA protocols.} Both \textit{Aura} and \textit{Clique} follow the basic idea of BFT-style consensus (cf. Fig.\ref{fig-poa-algm}), where only one authority leader determines the block with instant finality. The differences lie in two aspects: \textit{(i) the ways to elect this leader}; and \textit{(ii) the methods to agree on the block with peers}. In \textit{Aura}, the sealer rotation depends on the formula of $l=\frac{T}{duration} \mod N$, where $l$ represents the index of the leader, $T$ is the current UNIX time, $duration$ is the time interval between blocks and $N$ is the size of the committee (equal to the number of sealers). Once the leader is elected, the protocol launches a two-phase consensus: \textit{block proposal} and \textit{block acceptance}. The leader proposes a proposal (blocks containing a set of transactions) and broadcasts them to peers. Then, peer nodes (sealers) vote for this proposal. Only when the received responses exceed half of the involved sealers ($51\%$), this block is formally deemed to be confirmed. As in \textit{Clique}, the leader is rotated under the formula of $l=h \mod N$ where $h$ represents the block height. Such a way is inherently a \textit{round-robin} rotation, as $h$ increases linearly along with the chain growth. Compared to \textit{Aura}, the main difference is that \underline{multiple} sealers are allowed to propose blocks but restricted to propose a block every $N/2+1$ block. Here, the leader has a priority of proposing the block while non-leader sealers (but can still propose blocks) have to \underline{lag} their blocks with a random delay. The~\textit{difficulty} field of these non-leaders is set as $1$ whereas the leader's as $2$. To break ties\footnote{In blockchain systems, \textit{tie} means a temporary status where two competitive branches (subchains) have the \textit{same} length (or equiv. scores/weights/graphs) that results in difficulties in subchain selection by previous blocks.}, the chain with the highest scores will be selected as the valid chain (GHOST rule \cite{sompolinsky2015secure,sompolinsky2013accelerating}). Since a leader will be granted a higher score than other sealers, the block generated by this leader will be included in the final chain with a high probability.

Then, \textbf{we investigate their properties and assumptions.} According to its BFT-style design, PoA prioritizes \textit{consistency} over \textit{liveness} \cite{de2018pbft} (equiv. \textit{safety} and \textit{termination} in the context of BFT). Safety means all the honest sealers should have the same view of blocks at a specified height, while liveness emphasizes the honest blocks will eventually get included in the (longest) chain. To achieve safety, PoA relies on committee members (a.k.a. sealers) to offer deterministic finality. The proposed block, once agreed by the majority of sealers, will be instantly formatted as an unchangeable state without the possibility of being reversed. This provides faster confirmation and greater throughput. Liveness in PoA is achieved via both ongoing \textit{sealer rotation} and the GHOST rule (the highest-score/weightiest subchain wins). Sealer rotation makes the system still proceed in the case of single-point failures, and GHOST ensures that the chain will always grow under frequent forks. However, readers may easily omit an implicit assumption guaranteeing such properties: \textit{a majority of sealers (51\%) have to honestly behave in the synchronous network.} This assumption is very strict among existing distributed systems, which results in the lack of practicability in real-world applications. Despite this, we still find vulnerabilities under its strong assumption.

We observe that \textbf{PoA consensus protocols cannot guarantee \textit{fairness}.} Fairness focuses on the fair orderings, covering both block-level and transaction-level. The block-level means the block should be proposed in a proper turn. This requires their producers (\textit{a.k.a.} sealers) to faithfully take turns to mine blocks, without breaking the prearranged order. Transaction-level fairness refers to their fair orderings \cite{kelkar2020order,kelkar2021order} where transactions that have not yet been confirmed (pending status) confront the risk of being re-ordered by sealers for extra revenues (\textit{a.k.a.} miner extracted value, MEV \cite{daian2020flash}). These sealers may conduct arbitrary manipulation of the actual ordering of transactions by frontrunning, backrunning, or sandwitching \cite{zhou2021high} transactions inside their proposed blocks.  The threat of unfairness in PoA comes from its ambiguous statement of so-called \textit{honest behaviors}. The protocol merely assumes sealers should behave honestly, but in fact, they can act as well as \textit{profitably}. The task to order transactions in classical blockchain protocols is like \textit{dark forest} \cite{ethdkfrst} for miners. Similarly, PoA protocols follow such a design, in which transactions in the mempool \cite{daian2020flash} are intransparently ordered. This does great harm to users if the block producer intentionally manipulates their received transactions, causing a potentially huge amount of monetary loss for users who have honestly behaved in every step.

In this paper, \textbf{we explore the \textit{unfairness} issue in PoA protocols by initiating adversarial order manipulation attacks.} Unfairness is caused from two folds, and we accordingly initiate two types of profitable order manipulation strategies. Firstly, like miners in Ethereum~\cite{wood2014ethereum}, authority sealers in PoA-engined systems can arbitrarily sort transactions when sealing blocks, leaving an undiscovered area for adversarial manipulations by profit-pursuing sealers. This is a transaction-level frontrunning attack (denoted as the \textit{Type-I} attack) that is effective for all PoA implementations, as every honest block producer can insert her \textit{profitable} transaction to outpace others or delay peer transactions that are disadvantageous. Secondly, different from most permissionless consensus protocols (e.g., PoW) whose miners are randomly chosen from dynamic participants, PoA relies on a simplified BFT-style consensus with a static committee. The sealer rotation procedure depends on an open-sourced formula to select incoming block producers (more than one producer in \textit{Clique}). This brings threats because a profit-pursuing sealer can precisely predicate her turn to create blocks, and then forfeit the priority parameters (e.g., the field of \textit{delay time}) to occupy an advantageous position. 
Thus, we launch an attack (\textit{Type-II}) to frontrun the turn of privilege positions and further hold this position via broadcasting faked parameters (the very short \textit{delay time}). We implement these attacking strategies by slightly modifying the source code of Geth~\cite{go-clique} to simulate a profitable sealer. We apply both attacks in an isolated private PoA network (due to ethical reasons) and evaluate results in terms of effectiveness and success rates. Meanwhile, we submit the issue report to several projects and get their approval (cf. PR report \cite{hpbpr22}).

Furthermore, \textbf{we evaluate the potential loss of PoA markets affected by our attacks} and \textbf{provide in-depth discussions to mitigate the unfairness with our recommended countermeasures.} We review existing PoA-engined projects from Github\footnote{We search the keywords covering \textit{PoA} (2,392 repository results), \textit{Proof of Authority} (204), \textit{PoA Clique} (19), \textit{PoA Aura} (3) and \textit{authority round} (6). We remove overlapped or unrelated pages, obtaining the statistic results as below:  73\% of them are Clique-based systems (variants included), 18\% is Aura-based and the rest of 9\% represents others. [Data updated by Nov 11, 2021]} and select 10+ typical projects as our research focuses (see Tab.\ref{tab-impact}). The selected projects can cover most mainstream PoA implementations with their variations, where many projects directly apply PoA without any modifications (e.g., Ethereum Classic \cite{etc-clique}, Olecoin \cite{olecoin-clique}). We dive into these exemplified projects to explore the impact scope of our attacks. The \textit{Type-I} attack performs effective against \underline{all} types of PoA implementations, directly or indirectly influence up to $681,087$ million USD market cap [CoinMarketCap, Nov 2021]. In contrast, the \textit{Type-II} attack is merely effective on \underline{part of} \textit{Clique}-based projects. The reason lies in their different implementations of \texttt{verification} function. The function that checks both \textit{identity} and \textit{difficulty} for \underline{every} sealer who is sealing blocks can resist to our \textit{Type-II} attack, whereas the function checking merely \textit{difficulty} cannot prevent it, especially for non-leader block-producing sealers (\textit{difficulty} set as $1$,). This is because \textit{difficulty} is a fungible field in the checklist, requiring no further consensus procedures, while the \textit{identity} field has to be confirmed by all peer nodes. Our attack focuses on those non-leader sealers, manually shortening their \textit{delay time} (set to zero) for a priority of becoming dominant when the leader shuts down. Based on that, we provide our recommendations to mitigate the unfairness problem in PoA, avoiding the \textit{Type-II} attack as well as decreasing the potential loss. We provide three types of solutions. The first is to modify its core \texttt{verification} function, equipping identity checks. This type of solution applies direct modifications to source codes without additional complexity, but with sacrifices on system availability. The second approach introduces the physical randomness sampled from hardware. The third aims to make the turn of rotation unpredictable. This method requires on-top protocols or strong guarantees. A comprehensive discussion is attached in terms of their core benefits and drawbacks. In a nutshell, we summarize our contributions as follows.

\begin{itemize}
\item[-] We explore \textit{unfairness} in PoA protocols. We review 2,500+ \textit{in the wild} projects and select 10+ typical systems as our analysed targets. We dive into their source code and find the root cause that results in the unfairness issue.

\item[-] We initiate two types \textit{order manipulation} attacks against both transaction-level and block-level fairness of PoA protocols. We simulate our attack under the testnet of real-world projects (e.g., Ethereum Geth~\cite{go-clique}). Experiment results show that our attacking strategies can cause effective damages.

\item[-] We conduct a rigorous analysis of our attacks and evaluate potential loss in real-world markets. We compare the differences among the selected projects and discuss their potential attacking impacts in terms of on-site markets. Results indicate that the potentially affected market cap of these PoA projects can reach up to $681,087$ million USD.

\item[-] We give our recommended remedies against mentioned vulnerabilities. Our recommendations include various methods, covering code modification, hardware-assisted solutions, and crypto-based enhancement. We have further disclosed our results to some ongoing teams, referring to \cite{hpbpr22}.
\end{itemize}

\noindent\textbf{Paper Structure.} A high-level protocol design is presented in Section~\ref{sec-protoc}. The attacking strategies are introduced in Section ~\ref{sec-attack}, and are detailed in Section \ref{sec-typei} and Section \ref{sec-typeii}, respectively. A comprehensive discussion is presented in Section \ref{sec-discussion}. Our recommended fixes are provided in Section \ref{sec-recomd}. Related studies are discussed in Section \ref{sec-relatedwk}. Finally, Section \ref{sec-conclu} concludes our work. Appendix A-D provide supplementary details omitted in the mainbody.

\section{PoA Algorithms}
\label{sec-protoc}
This section presents two mainstream PoA implementations, \textit{Aura} and \textit{Clique}, by separately introducing their commons and operating mechanisms.

\subsection{PoA Model}
We abstract the PoA model from multiple related implementations. Despite their ways of completing consensus being different, we still find their commons in terms of participated roles, network assumptions, and adversary threat models.

\smallskip
\noindent\textbf{Roles.} PoA algorithms rely on a small group of authorities to make up the committee to perform consensus. Committee members are known as \textit{sealers}, who can package transactions and produce blocks. In each round, one of them will be selected as the \textit{leader} to propose the block and gather responses from others. To distinguish their roles, a sealer who becomes the leader is denoted as \textit{in-turn} sealer (\textit{difficulty} as 2), while others are \textit{out-of-turn} sealers as claimed in \cite{vincent20poa}. However, we clarify \textit{out-of-turn} sealers with more precise definition by adding a new term called the \textit{edge-turn} sealer, referring to a non-leader block-producing sealer (\textit{difficulty} as 1). The sealer who cannot produce any blocks are called \textit{out-of-turn} (\textit{difficulty} as 0). The in-turn sealer enjoys a priority of sealing blocks within her turns, whereas edge-turn sealers create blocks with a lagging time. 

\smallskip
\noindent\textbf{Assumptions.} PoA is designed for permissioned blockchain systems, where a limited size of sealers can seal blocks during the consensus procedures. We only put focus on communications among these committee nodes. The network is assumed to be partially synchronized, in which all the messages can be eventually delivered to their destinations with a maximum upper-bound delay of $\delta$. This means if one sealer sends a transaction to another sealer at the time $t$, the transaction will be delivered no later than the time $t+\delta$. We consider partial synchrony in our model due to its wide applicability towards most \textit{in the wild} projects. Few PoA implementations~\cite{parity} emphasize their total synchrony to guarantee their consistency and safety, but sacrifice the general adoption by real applications.

\smallskip
\noindent\textbf{Threat Model.} Previous attacks are generally based on negative assumptions such as the Byzantine failure in which a node acts arbitrarily: drop, delay, duplicate, or even fake messages. In contrast, our attack is based on a rational model where all the committee members behave honestly but can still pursue profits under predefined rules. We narrow down the scope of the utility function of these profits to simple monetary rewards, rather than any other type of virtual assets that are external to the system, like reputation mentioned in \cite{vincent20poa}. Therefore, in our attack, we assume two related profitable models. (i) an in-turn sealer (leader) can arbitrarily reorder the received transactions within her proposed block. This strategy does not break any honesty rules of the system (equiv. no modifications of any code). (ii) an edge-turn sealer (\textit{diff}=1) can continuously send falsified parameters (e.g., set lagging time as 0) to frontrun the advantageous position when the leader shuts down. This is a weaker promise than the former one, but still satisfies the legal rule because the sealer sends a valid number within the scope of protocol definitions.  Here, \textit{legal} means there are no checks towards \textit{the identity of edge-turn sealers} and \textit{delay time} in \texttt{verification}. Besides, the basic claim of $51\%$-honesty assumption for general PoA systems is kept as original in our attack. 

\subsection{PoA Implementations}
We provide two mainstream PoA implementations. In particular, we emphasize mechanisms from a block's proposal to verification.

\smallskip
\noindent\textbf{Aura Algorithm.} Aura (Authority Round) is implemented in Parity. The network in Aura is assumed to be synchronous, and all authorities are assumed to hold the same UNIX time $T$ (cf. Algm.\ref{alg-proposeaura} line 9). The leader of each step is deterministically calculated by duration (line 10) and the number of authorities $|sealers|$ (line 11), which is represented as $i = \frac{T}{duration}  \mod  |sealers|$. Each authority executes an infinite loop that periodically checks whether $i$ equals her position index. If it does, this authority will propose a block, and broadcast it to other authorities. If the received block is not produced by the current leader with correct difficulty $\textit{diff}$, it will be rejected. This is achieved by the verification algorithm (line 19-24). Otherwise, if the block is valid, the authorities will send received blocks to others. The leader is always expected to produce a block with correct difficulty, even if no transactions are available.

\begin{algorithm}
\caption{Block Propose and Block Verify in Aura}
\label{alg-proposeaura}
\begin{algorithmic}[1]

\Procedure{Queue}{$tx$}
\Comment{\textcolor{blue}{queue a transaction}}
\State $\textbf{if} (nonce\_check(tx) \wedge gas\_check(tx))$  \textbf{then}
\State $\qquad mempool \gets add(tx)$
\Comment{\textcolor{blue}{add tx to mempool}}
\State $\qquad broadcast(tx)$ \Comment{\textcolor{blue}{send the block}}
\EndProcedure

\State
\Procedure{Propose}{$sealer_i$}
\Comment{\textcolor{blue}{propose a block}}
\While{$(\mathsf{true})$} 
\State $\textit{T} = clock-time()$
\State $step \gets T/duration$
\State $\textbf{if} (i \in sealers \wedge step \mod |sealers| = i)$  \textbf{then}
\State $\qquad TXs \gets mempool.pop()$ 
\Comment{\textcolor{red}{attack point-I}}
\State $\qquad block \gets sign(TXs)$ \Comment{\textcolor{blue}{in-order sealing}} 
\State $\qquad broadcast(block)$ \Comment{\textcolor{blue}{send the block}}
\State $sleep(step-duration)$
\Comment{\textcolor{red}{attack point-II}}
\EndWhile\label{euclidendwhile}
\EndProcedure


\State
\Procedure{Verify}{$sealer_i$}
\Comment{\textcolor{blue}{verify a block}}
\State \textit{diff} $=$ \textit{calculate\_difficulty(header.number, sealer)}
\State \textbf{if} ($\textit{diff}\,\, == \textit{header.difficulty}$) \textbf{then} \Comment{\textcolor{blue}{block is valid}}
\State $\qquad \textbf{return}$ $\mathsf{true}$
\State $\textbf{else}$ \textbf{then}
\State $\qquad \textbf{return}$ $\mathsf{false}$ \Comment{\textcolor{blue}{block is invalid}}
\EndProcedure

\end{algorithmic}
\end{algorithm}

\smallskip
\noindent\textbf{Clique Algorithm.} \textit{Clique} is a PoA algorithm that was first implemented by Geth~\cite{go-clique}. Similar to the Aura algorithm, in each round, a new block is proposed by the current leader. However, a leader is selected by combining the block number and the number of authorities rather than the UNIX time (cf. Algm.\ref{alg-proposeclique} line 11). The main difference between \textit{Clique} and other algorithms is that the selected leader is not exclusive in each block period, while other authorities are still allowed to propose blocks with a random-delayed time (Algm.\ref{alg-proposeclique} line 16). The parallel block-producing mechanism effectively prevents a single Byzantine authority from havoc on the whole network. Forks may frequently occur where a block is proposed by more than one authority. \textit{Clique} solves this issue by two approaches. Firstly, a random delay time with \textit{wiggle} must be added to a new block that is created by a non-leader block-producing sealer (Algm.\ref{alg-proposeclique} line 15). Thus, the block produced by a leader has advantageous possibilities of being accepted by peer sealers. Secondly, an accumulation scoring mechanism and the GHOST protocol (or longest chain rule, depending on specific implementations) have been applied to resolve forks. In particular, a leader's block will be granted higher scores than peers. Based on that, GHOST \cite{sompolinsky2015secure} guarantees that such a block will be eventually accepted by a majority of authorities. Here, we emphasize that in each round, the number of authorities, including the leader and authorities who can propose delayed blocks (in/edge-turn), is limited to $|sealers| - (\frac{|sealers|}{2} + 1)$ (Algm.\ref{alg-proposeclique} line 37-42, with associated source code of Geth at ~\cite{go-clique} line 477-483). The design provides high availability and simultaneously reduces network redundancy for the system, guaranteeing the system can still operate when the in-turn sealer and followed edge-turn sealers (not all) shut down.

\begin{algorithm}
\caption{Block Propose and Block Verify in Clique}
\label{alg-proposeclique}
\begin{algorithmic}[1]

\Procedure{Queue}{$tx$}
\Comment{\textcolor{blue}{queue a transaction}}
\State $\textbf{if}\,  (nonce\_check(tx) \wedge gas\_check(tx))$  \textbf{then}
\State $\qquad mempool \gets add(tx)$ \Comment{\textcolor{blue}{add tx to mempool}}
\State $\qquad broadcast(tx)$ \Comment{\textcolor{blue}{send the block}}
\EndProcedure

\State
\Procedure{Propose}{$sealer_i$} \Comment{\textcolor{blue}{propose a
block}} 
\While{$(\mathsf{true})$} 
\State $n \gets lastblock.number$
\State $\textbf{wait until} \neg sign\_recently(sealer_i,n)$
\Comment{\textcolor{blue}{wait to seal}}
\State $\textbf{if}\, (n+1) \mod |sealers| = i$ \textbf{then}
\State $\qquad weight = 2$ \Comment{\textcolor{blue}{in-turn sealer}} 
\State $\textbf{else}$
\State $\qquad delay = block.time - now$
\State $\qquad wiggle = (\frac{|sealers|}{2} +1) * 500ms$ \Comment{\textcolor{blue}{wiggle time}}
\State $\qquad delay = rand(delay + wiggle)$ \Comment{\textcolor{red}{attack point-II}}
\State $\qquad sleep(delay)$ \Comment{\textcolor{blue}{sleep for a while}}
\State $ TXs \gets mempool.pop()$ 
\Comment{\textcolor{red}{attack point-I}}
\State $ block \gets sign(TXs,wiggle)$ \Comment{\textcolor{blue}{seal a block}}
\State $broadcast(block)$ \Comment{\textcolor{blue}{send the block}}
\EndWhile\label{euclidendwhile}

\EndProcedure

\State
\Procedure{Verify}{$sealer_i$}
\Comment{\textcolor{blue}{verify a
block}}
\State $\neg sign\_recently(sealer_i,block.number)$
\State \textit{diff} $=$ \textit{header.difficulty}
\State $\textbf{if}\,  (\textit{diff}\,== nil)$ \textbf{then} \Comment{\textcolor{blue}{not nil}}
\State $\qquad \textbf{return}$  $\mathsf{false}$ 
\State $\textbf{else if}\,  (\textit{diff} \,\, !=  1 \, \&\& \,\textit{diff} \,\,!= 2)$ \textbf{then} \Comment{\textcolor{blue}{1 or 2}}
\State $\qquad \textbf{return}$ $\mathsf{false}$
\State $\textbf{else}$ \textbf{then}
\State $\qquad \textbf{return}$ $\mathsf{true}$ \Comment{\textcolor{blue}{block is valid}}
\EndProcedure

\State
\Procedure{sign\_recently}{$sealer, number$}
\For{$(\mathsf{i < |recents|})$} \Comment{\textcolor{blue}{amongst recent sealer, wait}}
\State $\textbf{if}\,(\mathsf{sealer == recents[i]})$ 
\State $\qquad limit = \frac{|sealers|}{2} +1$
\State $\qquad \textbf{if}\,  (number < limit || i > number-limit)$ \textbf{then}
\State $\qquad \qquad \textbf{return}$  $\mathsf{false}$ 
\State $\textbf{return}$ $\mathsf{true}$ 
\EndFor
\EndProcedure
\end{algorithmic}
\end{algorithm}

\section{The Ordering Attack}
\label{sec-attack}

This section introduces our attacks, covering both \textit{Type-I} (frontrunning attack in transaction-level) and \textit{Type-II} (break the proper turn of enrolled sealers) attacks.  We present a bird view of our attacks. Instances towards \textit{Aura} and \textit{Clique} are discussed in Section \ref{sec-typei} and Section \ref{sec-typeii}, respectively.

\subsection{Breaking Fairness of Transaction Ordering}

Transactions once sent by users will firstly arrive in a pending pool \cite{daian2020flash}, waiting for the miners' pickup. In most cases, miners conduct transactions according to a set of priorities such as paid fees (e.g., Ethereum \cite{wood2014ethereum}), timestamp, randomness or combinations of them (as discussed in \cite{wang2020sok}). A user can facilitate her transaction by increasing the gas fee if she has lagged out of peers. It is an eligible way for regular users because the rules are known and agreed upon by them in advance. However, that fact is that miners have advantageous opportunities to earn more revenues by inserting/reordering/delaying transactions in addition to normal block rewards, as they can perform arbitrage strategies without being monitored. Based on that, fairness, in terms of transaction ordering, requires that miners (sealers) to \textit{execute received transactions in a proper turn} according to their predefined rules. Alternatively, the order of transactions is sorted merely by their intrinsic properties rather than external manipulation.

\begin{figure}
\subfigure[\textbf{The \textit{Type-I} Attack:} The attack aims to break the transaction ordering fairness by inserting profitable transactions $Tx_{s}$ (by ways of displacement, insertion, or suppression, as shown in red dotted boxes). The attack happens within the same block when an in-turn sealer produces this block. It is a transaction-level attack that can be applied to any PoA implementations only if they still rely on a block producer (or miner).]{
\begin{minipage}[b]{\linewidth}
\centering
\includegraphics[width=0.8\linewidth]{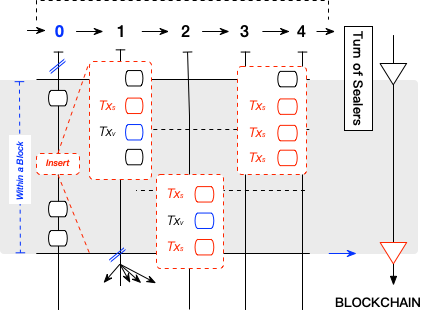}
\end{minipage}
}
\subfigure[\textbf{The \textit{Type-II} Attack:} The attack targets to break the regular turn of the sealer rotation. A profitable (edge-turn) sealer may frontrun other (edge-turn) sealers when the leader shuts down by sending a falsified \textit{delay time}. For a fixed attacker, she can periodically send the falsified value to occupy the position in discrete rounds out of her turn. For dynamic attackers, they can seize an advantageous position in successive rounds. The attack is feasible in several PoA implementations as the associated \texttt{Verification} function merely checks \textit{difficulty} (pass if $1$), instead of their identities. ]{
\begin{minipage}[b]{\linewidth}
\centering
\includegraphics[width=0.8\linewidth]{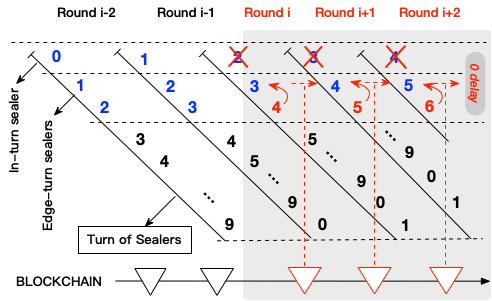}
\end{minipage}
}
\caption{The Ordering Attacks against PoA Protocols}
\label{fig-attack}
\end{figure}

The \textit{Type-I} attack aims to enforce the manipulation of transaction ordering (see Fig.\ref{fig-attack}.a). This attack typically needs to find a profitable transaction as the target. We assume the victim transaction as \texttt{Tx$_v$}, and the frontrunning transaction as \texttt{Tx$_s$}. As discussed in \cite{torres2021frontrunner}, the manipulation covers three types of activities, namely \textit{displacement}, \textit{insertion} and \textit{suppression}. \textit{Displacement} means the sealer's transaction \texttt{Tx$_{s}$} is sorted before the victim's \texttt{Tx$_v$}. \textit{Insertion} includes two sealers' transactions \texttt{Tx$_{s1}$} and \texttt{Tx$_{s2}$}, where \texttt{Tx$_{s1}$} is ordered before \texttt{Tx$_v$} and \texttt{Tx$_{s2}$} is after \texttt{Tx$_v$} (also known as the sandwich attack). \textit{Suppression} indicates that the attacker sends a set of \texttt{Tx$_{s}$} to fill up the block by reaching the maximum of gas limits. In this way, \texttt{Tx$_v$} is suppressed and cannot be included in the next block. We apply the first strategy and second strategy to our PoA model, and skip the third strategy, as this type of attack is generally launched by extra assumptions. Exceeding the gas limit will cost a lot of fees, greatly reducing the miner's revenue. We provide our detailed attacking procedures in Section \ref{sec-typei}.

\subsection{Breaking Fairness of Election Turn}

Compared with transaction-level ordering, another important aspect of \textit{fairness} is to keep the proper order of sealers' positions, especially in BFT-style consensus that relies on powerful authorities. The PoA algorithm, as a variant of BFT protocols, follows similar principles. The identities of sealers in PoA are visible to anyone, resulting in the risk of being manipulated by internal (profitable) parties. An edge-turn sealer may spare no effort to pursue profits by ways of frontrunning the position to produce blocks when the leader shuts down. Even worse, some PoA implementations (e.g., \textit{Clique}) provide probabilistic finality where the chain grows with inevitable forks. The faster a chain grows, the more forks it produces. Any unpredictable events such as delayed blocks or malicious occupation may lead to system failure. Therefore, making sealers sit in correct positions during the rotation can reduce negative impacts.

Our \textit{Type-II} attack targets to break the eligible order of sealer rotation. The attack is to seize a leading position among all edge-turn sealers. Assume that totally nine sealers participate in the network, one of them is an in-turn sealer $N_0$, two are edge-turn sealers $N_3, N_4$ and others are our-of-turn sealers $N_5 - N_8$ with the turn rotation of $N_0 \rightarrow N_1 \rightarrow ... \rightarrow N_8$ (cf. Algm.\ref{alg-proposeaura} line 10-11 and Algm.\ref{alg-proposeclique} line 11). Here, we suppose that the attack starts at the round $i$ and $N_4$ in this round is a profitable edge-turn sealer who attempts to frontrun the turn of $N_3$ (see Fig.\ref{fig-attack}.b). Similar attacks of profitable sealers $N_5$ in round $i+1$ and $N_6$ in round $i+2$, are also applied against benign sealers $N_4$ and $N_5$, respectively.

Both \textit{Aura} and \textit{Clique} create an eligible \textit{difficulty} of either $2$ for in-turn sealer or $1$ for edge-turn sealers in \texttt{Block Propose}. The block with $2$ will outpace peers due to its highest priority of the weight (Algm.\ref{alg-proposeclique} line 12). Our attack performs \underline{ineffective} in this situation because an in-turn sealer must be set with $2$ (a strict verification step, Algm.2 line 37-42). Therefore, we put our focus on edge-turn sealers who can also generate blocks when the leader fails for some reason (e.g., being compromised). The key of our attack is to frontrun the position of others sitting in front. Since all edge-turn sealers equivalently hold the \textit{difficulty} of $1$, peer nodes will pass the first-received block without having to distinguish who has sent them. Our attack exactly leverages the unexplored area in the ambiguous verification towards edge-turn sealers: \textit{peer nodes merely check whether the difficulty field is legal but omit the check on their origins}. Based on that, our modifications of the source code are constrained merely to the \textit{delay time} of edge-turn sealers (Algm.\ref{alg-proposeclique} line 17), without breaking any code rules. Our modifications pose negligible impacts to peer nodes. However, due to significant differences between \textit{Aura} and \textit{Clique} in their \texttt{Block Verify}, our \textit{Type-II} attack against them will generate very different results, as discussed in Section \ref{sec-typeii}.

\smallskip
\noindent\textbf{Experiment Configuration.} To test the feasibility of our attacks, we clone the source code from Geth \cite{geth} and apply our attacking logic to it. Then, we run a private PoA network in an isolation environment. Our experiments run on Centos $7$ $x86\_64$ at a virtual private server (VPS) located in California, US, with hardware configurations of 4GB RAM, 80G disk storage, and 3000GB bandwidth capacity. We create several sets of virtual nodes in this VPS to simulate independent consensus sealers. Then, we start our attacks and investigation. The details are discussed in the following sections.

\section{Breaking Transaction Orders}
\label{sec-typei}

In this section, we show how to apply the \textit{Type-I} attack against \textit{Aura} and \textit{Clique}. We provide the experiment configurations, logic, results, and attacking impacts, respectively.

\smallskip
\noindent\textbf{Experimental Logic.}
A transaction is required to experience four stages from user's submission to final confirmation by sealers: (i) A user initiates a transaction with the signature signed by her private key, and then sends the transaction to blockchain nodes. (ii) The blockchain node verifies this transaction and adds it to the local transaction pool. Then, the node broadcasts the transaction to peers. The peer nodes perform the same procedures by validating and moving transactions into their transaction pools. (iii) As a specific type of node, the blockchain sealer orders received transactions and packages them into a block. Sealers have the flexibility to decide transaction orders. Taken Geth \cite{go-clique} as an example, transactions are ordered by gas price and nonce value. Generally, the higher the gas fee is provided, the faster a transaction enters the block. (iv) As mining nodes receive the new block from their peers, they will re-execute all transactions inside the block. When a majority of nodes receive the same block, and re-execute included transactions, the block is deemed as confirmed and becomes immutable (together with its contained transactions). This ensures that all blockchain nodes have agreed on the same history of transactions.

\subsection{Running the Attack-I}
\smallskip
\noindent\textbf{Attack Simulation.} In our experiments, the attacker is assumed to be a profitable sealer with the intention to manipulate transactions by frontrunning and inserting. To be simple, we denote all these profitable transactions as frontrunning transactions for generality. The attacker is empowered with capabilities to find arbitrage opportunities and accordingly generate a frontrunning transaction (refer to \cite{daian2020flash,torres2021frontrunner} for ways to create such transactions). Then, the attacker inserts her own transaction before/after the user's transaction during the procedure of sealing a block. 

To launch the attack, an attacker actively checks arbitrage opportunities. Afterward, the attacker issues one transaction $Tx_s$ and attempts to insert $Tx_s$ at an advantageous position in the mempool (equiv. a fronted position). Here, the mempool requires slight modification from the code aspects. Then, it seals a block and broadcasts to other peers. The \textit{Type-I} attack is considered successful if a block contains a list of transactions that the users' transaction is ordered after $Tx_s$. It means that we have successfully allocated the attacker's frontrunning transaction before a targeted transaction in the queue. Then, this block is successfully verified and confirmed by the majority of honest miners. To simulate this attack, we present a simple way to apply the frontrunning attack by sending a series of transactions $Tx_1, Tx_2,..., Tx_n$ to a miner who will generate a frontrunning transaction $Tx_s$. To cover accidents and achieve a fair testing result, we repeat tests for six times. 

\begin{table}[ht]
\centering
\caption{Attack-I Experiment Results}
\label{tab-experiment1}
\begin{threeparttable}
\resizebox{0.9\columnwidth}{!}{%
\begin{tabular}{c>{\columncolor{green!20}} c c>{\columncolor{green!20}} c c>{\columncolor{green!20}} c c>{\columncolor{green!20}} c c c}
   \midrule
   \textit{\textbf{Set}}  & \multicolumn{9}{c}{\textit{\textbf{Time Slots}}}\\
   \midrule
   \textbf{Experiment 1} \qquad & & $Tx_1$ & & $Tx_2$ &  & $Tx_3$ & & $Tx_4$ & \multirow{6}{*}{...} \\
   
   \textbf{Experiment 2} \qquad & $Tx_s$ & $Tx_1$ &  & $Tx_2$ &  & $Tx_3$ & & $Tx_4$ &\\
   \textbf{Experiment 3} \qquad & & $Tx_1$ & $Tx_s$ & $Tx_2$ &  & $Tx_3$ & & $Tx_4$ & \\
   \textbf{Experiment 4} \qquad & & $Tx_1$ &  & $Tx_2$ &  $Tx_s$ & $Tx_3$ & & $Tx_4$ &\\
   \textbf{Experiment 5} \qquad & & $Tx_1$ &  & $Tx_2$ &   & $Tx_3$ & $Tx_s$ & $Tx_4$ &\\
   \textbf{Experiment 6} \qquad & & $Tx_2$ &  $Tx_s$ & $Tx_4$ & $Tx_s$  & $Tx_3$ & $Tx_s$ & $Tx_1$ & \\
   \midrule
\end{tabular}
}
  \begin{tablenotes}
       \footnotesize
       \item Transactions with \colorbox{green!20}{\color{black}Green} background are profitable transactions.
     \end{tablenotes}
  \end{threeparttable}   
\end{table}%

\subsection{Attacking Result Analysis} 

In the primary version of experiments, we conduct tests on a single node to check the attacking feasibility. Here, both \textit{Aura} and \textit{Clique} show a very similar result (see Tab.\ref{tab-experiment1}). \textit{Experiment 1} shows the original order of the transaction list ($Tx_1$,$Tx_2$,$Tx_3$,$Tx_4$), while \textit{Experiments 2-6} show the results under our attacks. From these results, we observe that the frontrunning transaction $Tx_s$ can be arbitrarily inserted to any position between these transactions. Even worse, the transaction list ($Tx_1$,$Tx_2$,$Tx_3$,$Tx_4$) can be re-arranged in an arbitrary order ($Tx_2$,$Tx_4$,$Tx_3$,$Tx_1$) by a profitable in-turn sealer. 

Further, we extend our attack to multiple nodes (cf. the blue line \underline{without} Attack-II of each sub-figure in Fig.\ref{fig-result}). Our applied attacks towards \textit{Aura} and \textit{Clique} both perform effective. In \textit{Type-I} case, all sealers behave honestly but can pursue more revenues. Firstly, we fix the number of sealers as 9. The success rate of this attack is approximate $11\%$ (Fig.\ref{fig-result}.b/d), proved by its absolute number of victim transactions (Fig.\ref{fig-result}.a/c). Then, we investigate the situations under dynamic committee where the size of committee ranges as \{3,9,18,27\}. The result of associated success rates is inversely proportional to participated nodes (Fig.\ref{fig-result}.e/f/g/h). This is easy to understand when we notice a fact: \underline{only} the in-turn sealer can successfully arbitrarily insert her frontrunning transaction. Otherwise, the proposed block by an edge-turn sealer cannot out-complete an in-turn sealer due to their differences in \textit{difficulty}. Thus, the more participants get involved in the consensus, the less opportunity an attacker's block could be accepted. Our attack towards \textit{Aura} and towards \textit{Clique} output the same results in terms of both victim transactions and their success rates, indicating that \textit{Type-I} is a general attack that can be applied to \underline{all} PoA implementations based on the current knowledge. 

\smallskip
\noindent\textbf{Root cause.}  Typically, the disorder of transactions comes twofold. Firstly, even if an honest sealer follows a fixed order rule, the final result may still be disordered. In a peer-to-peer network, the same set of transactions may arrive at different sealers with entirely different orders due to the network delay. This situation happens with a non-negligible probability before the final decision of the consensus agreement. Inconsistent sequence status may exist for a long time, which leaves enough time for conducting profitable activities by in-turn sealers. Secondly, as mentioned earlier, the sealer can decide how to sort transactions. When a profitable sealer is mining, she can re-arrange transactions in an arbitrary order within her block to obtain the best interests. Meanwhile, other honest sealers cannot distinguish who is the profitable sealer and whether she has done evil.

In the current version of \textit{Aura} and \textit{Clique}, their implementations lack of a verification mechanism of transaction orders. In \textit{Aura}, the order of transactions depends on the preference or discretion of a sealer (see Algm.\ref{alg-proposeaura} line 12). Without any order-check mechanism in the verification procedure, a sealer can always pursue the highest transaction fees by rearranging the order of received transactions. Similarly, the implementation of \textit{Clique} shares the same transaction order mechanism (Algm.\ref{alg-proposeclique} line 16). Despite existing PoA algorithms prescribing specific rules for consensus, guarantees are absent on the relationship between the sequence of incoming transactions and their final order published in the ledger. Thus, PoA algorithms cannot promise transaction-order-fairness \cite{kelkar2020order}.

\section{Breaking Sealer Turns}
\label{sec-typeii}

\smallskip
\noindent\textbf{Experimental Logic.} PoA consensus algorithm relies on a rotation schema, where sealers take turns to mine blocks in different rounds based on the time and block number. In PoA, the time is divided into steps, and one of the authorities is elected as the leader to produce a blockchain in each step. The leader, an in-turn sealer, can generate a block exactly in her turn, while an edge-turn sealer can also produce blocks. By the design and assumption of PoA algorithms, these sealers follow a fixed formula (Algm.\ref{alg-proposeaura} line 11 and Algm.\ref{alg-proposeclique} line 11) to calculate their turns. An in-turn sealer can propose a block with a high priority of the weight. Edge-turn sealers with the secondary priority have to produce blocks when the leader fails. In this series of experiments, we test the potentiality of extracting extra benefits of those edge-turn sealers. In every \textit{single}\footnote{The term \textit{single} indicates that, if we merely test the \textit{Type-II} attack, an in-turn sealer should immediately keep sleeping to ensure the valid actions of edge-turn sealers. But when adding the \textit{Type-I} attack, things become different. The \textit{Type-I} attack relies on the manipulation of authorised sealers, no matter she is a legal in-turn sealer, or a profitable edge-turn sealer. Therefore, in case of testing the \textit{hybrid} \textit{Type-I}\&\textit{Type-II} attack, the original legal in-turn sealer should keep alive at in initial stage.}  \textit{Type-II} attack, we sleep the legal in-turn sealer to activate competitive edge-turn sealers. Then, we separately adjust the number of sent transactions/blocks and the size of participated committee to simulate different real scenarios. In each \textit{hybrid} attack (\textit{Type-I}+\textit{Type-II}), we keep the legal in-turn sealer alive and conduct the same experimental procedures for evaluation. Details are presented as follows.

\subsection{Running the Attack-II}
\noindent\textbf{Attack Simulation.} To enable the capability of frontrunning blocks, we allow an attacker to tamper with a mining client (slightly modify the source code at the \textit{delay time} field\footnote{Revised source code available at: \url{https://github.com/poa-research/attacksimulator}.}). The modified client can keep sealing a block in every round without considering the condition of block generation (always sending $0$-time delayed blocks). Notably, only the attacker's code needs to change, while all the peer nodes are still supposed to be benign without modifications. Details of our modifications in \textit{Aura} and \textit{Clique} are shown as below. (i) Change the UNIX time of the local client (Algm.\ref{alg-proposeaura} line 9) or delete the conditional expressions (Algm.\ref{alg-proposeaura} line 11 ) to ensure the profitable sealer can always seal a block in each round. (ii) Delete all the conditions of producing a block in the non-leader setting (Algm.\ref{alg-proposeclique} line 14-17 ), and allow an edge-turn sealer to mine instantly in every round with the score $1$. Afterward, the profitable sealer re-compiles the source code to obtain a heterogeneous client. Then, both \textit{Aura} and \textit{Clique} conduct the following attack simulations.

\begin{figure*}
\centering
\includegraphics[width=1\linewidth]{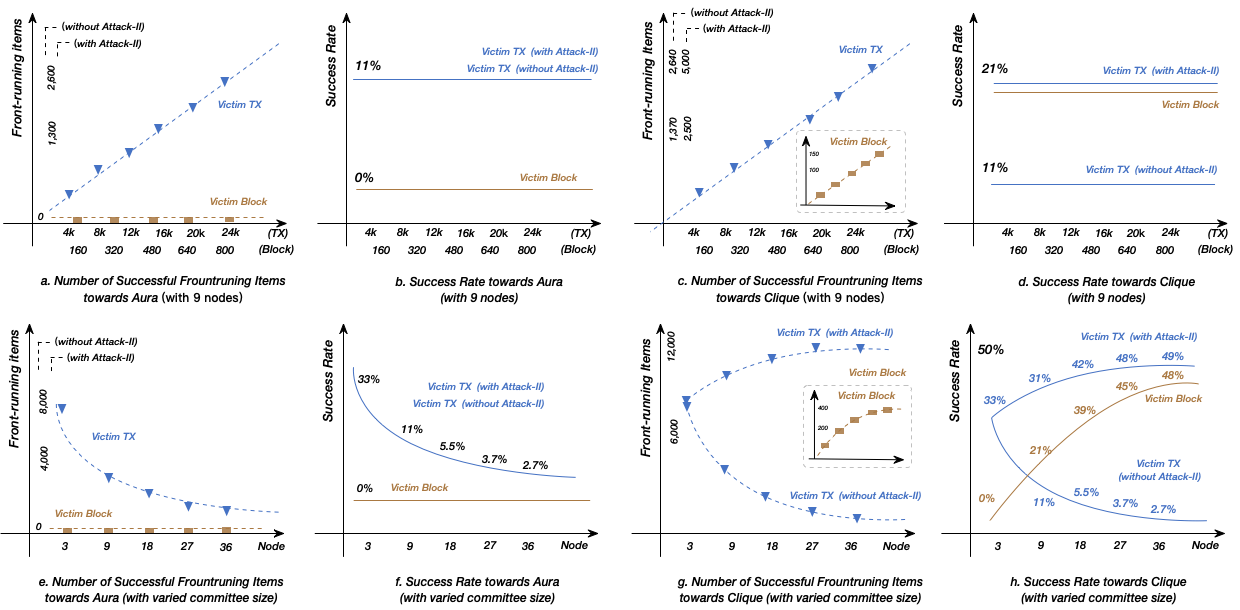} 
\caption{\textit{The Order Manipulation Attacks}: We launch two sets of experiments. In the first set [Fig.a-d], we set the committee size as 9, and the $x$-axis implies the total sent transactions (upper) or blocks (bottom). We count the successful frontrunning transactions/blocks (equiv. victim TXs/blocks) towards \textit{Aura} [Fig.a] and \textit{Clique} [Fig.c]. Based on that, we calculate the success rates of both transaction-level attack [Fig.b] and block-level attack [Fig.d]. In the second set [Fig.e-h], we set a total upper bound of sent transactions as $18,000$, and vary the committee size in $x$-axis. Similarly, we conduct attacks against \textit{Aura} [Fig.e/f] and \textit{Clique} [Fig.g/h].  The results demonstrate that the \textit{Type-I} attack is effective on both PoA implementations, whereas \textit{Type-I} merely perform effective in \textit{Clique}. Notably, the block produce rate is $3$s per node and the transaction sending rate is 100 TPS per node. Results of \textit{Type-I} attacks are presented in blue while \textit{Type-II} in brown.}
\label{fig-result}
\end{figure*}

\smallskip
The first experiment mainly evaluates victim transactions (with the additional \textit{Type-II} attack compared to the previous section) and blocks along with an increasing number of total transactions/blocks (marked by $x$-axis in Fig.\ref{fig-result}, which is based on a time-increased line). Here, victim blocks represent the blocks being frontrun by profitable edge-turn sealers, whereas the in-turn blocks \textbf{cannot} be deemed as the victim blocks.

To achieve this simulation, we set the $9$ mining nodes and $1$ boot node in our network. The mining nodes can synchronize transactions and block data, and merge transactions into blocks. All those mining nodes, in our attack experiments, are granted with the ability to become edge-turn sealers with $\textit{diff}=1$, but not in-turn sealers with $\textit{diff}=1$ if out of their turns. Also, every legal in-turn sealer is intentionally set to sleep. This is to simulate the case of an in-turn sealer's shutting down. As a result, edge-turn sealers stay active when solely testing the \textit{Type-II} attack. In contrast, the boot node is only used for node discovery and does not need to synchronize chain data. All instances, including the malicious client, are connecting with each other. Then, we set the block generation rate at $3$ seconds, which indicates that each block is generated at an average time of $3$ seconds. An attacker is supposed to be a profitable edge-turn sealer with intentions to frontrun blocks, hidden in $9$ mining nodes. The attacker then aims to produce a valid block out of her turn. To start the attack, the malicious sealer runs the modified client under the help of the boot node for joining the network and starting to mine.  We allow all sealers to mine and users to continuously send transactions to any sealer at a constant rate ($10$ tx/s) within a fixed period ($40$ minutes). The \textit{Type-II} attack is considered successful if a profitable (edge-turn) sealer can seal blocks before peered sealers.

The second experiment attempts to evaluate the relationship between the number of mining nodes and victim items. Again, we set the block generation rate at $3$ seconds. Since PoA is more suitable for permissioned blockchain systems \cite{de2018pbft}, we limit the size of mining nodes to an upper bound of $36$. More specifically, we set $3$ mining nodes and 1 boot node in our network. Then, the block generation rate is $3$ seconds per block per node, while the transaction generation rate is $10$ TPS per node. We adjust the participated mining nodes as \{$9$, $18$, $27$, $36$\} at the cumulative time of $40$ minutes. The experiment will show the trend between the probability of sealing a block and the number of mining nodes (sealers).

\subsection{Attacking Result Analysis.} 

In this part, we analyze the attacking results from two sides. Firstly, we fix the size of committee with $9$ sealers (cf. Fig.\ref{fig-result}.a-d). We test the successful frontrunning transactions and corresponding success rates towards both \textit{Aura} and \textit{Clique}. Then, we put the focus on the dynamic committee by increasing participated sealers (Fig.\ref{fig-result}.e-h). Here, we provide detailed discussions.

Intuitively, in the first group of experiments, $800$ blocks are produced in total. The created blocks will be uniformly distributed in $9$ mining nodes (or sealers), meaning that each node will generate approximately $88$ blocks. In \textit{Aura}, the results of successful frontrunning blocks remain zero, satisfying our expectation: Although a profitable sealer can generate $90$ blocks every round, however, her proposal \textbf{cannot} pass the verification (cf. Algm.\ref{alg-proposeaura} line 19-24) because Aura does not rely on the mechanism of \textit{difficulty} as it in Clique. The adversary has negligible possibilities to frontrun blocks (Fig.\ref{fig-result}.a). Accordingly, the success rate turns to be zero as well (brown line in Fig.\ref{fig-result}.b). Then, if we enforce both \textit{Type-II} and \textit{Type-I} attack, the result of victim transactions and success rate keep consistent with results in the previous section, where each block should contain approximately $88$, with a rate of $11\%$ ($11\% \approx \frac{1}{9}$, blue line in Fig.\ref{fig-result}.a/b). On the contrary, in \textit{Clique}, a profitable sealer can re-order any transactions since she can frontrun all blocks produced by edge-turn sealers in front of her under the absence of the leader (brown line in Fig.\ref{fig-result}.c/d). The indicates \textit{Clique} is vulnerable to our attacks, especially for \textit{Type-II}. Intuitively, victim transactions (blue line in Fig.\ref{fig-result}.c/d) follow the same trend with victim blocks as every edge-turn sealer, once become the leader by frontrunning, can dominate all included transactions within her blocks.

In the second group of the experiment of \textit{Aura}, we observe that the number of victim blocks stays zero at all the time (brown line in Fig.\ref{fig-result}.e/f), which means that \textit{Type-II} attack is ineffective to \textit{Aura} no matter how many nodes are involved. A profitable sealer has no opportunities to conduct the block-level frontrunning attacks out of her turns. As for the victim transactions, the number of transactions mined by the profitable sealer and its associated rate always exist but start to decrease along with increased nodes (blue line in Fig.\ref{fig-result}.e/f). This is easy to understand: Suppose that there is a profitable sealer in the committee, even though she cannot frontrun the turn of others, she can still send frontrunning transactions within her legal turn. Similarly, the number of these frontrunning transactions declines when the number of participated sealers increases. This confirms our predictions that transactions are uniformly distributed in different blocks, decreasing the frontrunning blocks. In \textit{Clique}, adding the \textit{Type-II} attack will both affect the number of victim transactions and victim blocks in a positive relation (Fig.\ref{fig-result}.g/h). An interesting point is the success rate of victim transactions (adding Attack-II) is approximately equal to the sum of another two (victim block\&victim Tx without Attack-II in Fig.\ref{fig-result}.h). This is because the count of victim transactions includes both sources from the in-turn sealer (her legal turn) and edge-turn sealers who successfully conduct frontrunning. In this case, a profitable sealer can always seize edge-turn sealers' opportunities to seal the block with included transactions. PoA is thereby vulnerable to both \textit{Type-I} and \textit{Type-II} attacks.

\smallskip
\noindent\textbf{Root cause.} We have dived into the source code of both \textit{Aura} and \textit{Clique}, abstracted their operating workflow, and found the root cause resulting in the \textit{Type-II} attack. The attack is ineffective to \textit{Aura} whose consensus is designed from BFT protocols, totally different from \textit{Clique}. We skip the discussion of \textit{Aura}. The reason that the \textit{Type-II} attack is effective to \textit{Clique} lies in \textit{the incomplete design of identity-verification mechanism towards edge-turn sealers}, which, unfortunately, indicates that the identity-verification towards an in-turn sealer applied to real implementations is valid, whereas similar checks towards edge-turn sealers are invalid, or at least, incomplete. It is clear that both the in-turn sealer (high priority) and edge-turn sealers (secondary priority) can produce blocks. When the in-turn sealer fails, edge-turn sealers start to become important. However, even though the protocol has checked\&passed $\textit{diff}=1$ for those edge-turn sealers, a serious fact has been ignored: \underline{multiple} active sealers hold $\textit{diff}=1$. This is the most significant difference compared to the case with an in-turn sealer because only \underline{one} eligible candidate holds $\textit{diff}=2$ in each round. Meanwhile, all edge-turn sealers are required to add a random delay during their block production. When the only one in-turn sealer fails, an adversary can easily break the normal turns of active edge-turn sealers by shortening the time delay (cf. Fig.\ref{fig-attack}). Therefore, the lack of a complete identity-verification mechanism towards edge-turn sealers enables profitable ones to break the fairness of normal rotation turns. 

\smallskip
Beyond the fairness issues, we also notice that \textit{our order manipulation attacks can result in a decrease in performance.} Here, the performance represents all valid transactions processed by benign block-producing sealers (excluding profitable sealers). A profitable sealer can rationally create many empty blocks for fast delivery, outpacing heavy transactions sent by peers. This will significantly hinder the normal procedure of packing real transactions from users. A user may have the feeling that she has waited for a long time whereas the transaction has still not been confirmed. In our predictions, a rational sealer will conduct such activities with high probability even if extra revenues are not as much as expected.

\section{Analysis and Discussion}
\label{sec-discussion}
This section discusses the impact of our attacks from the view of blockchain properties, comparisons with common attack vectors, attacking scope, and potential monetary loss. 

\subsection{Property Analysis}

We consider three orthogonal properties\footnote{In the context of Byzantine style consensus, consistency is cited by \textit{safety}, sometimes coming with \textit{validity}. Liveness is known as \textit{termination}. We regard these terms as the same meaning in this work.}  (including consistency, liveness~\cite{wang2020sok} and fairness~\cite{kelkar2020order}) that closely relate to the system security. Consistency is to ensure all honest nodes in the network decide on the same state while liveness guarantees all correct nodes eventually have decisions. These two properties refer to the intrinsic designs of a system that have been studied for decades. In contrast, fairness~\cite{kelkar2020order}, as a new proposed property, has been overlooked for a long time due to the particularity of the miner mechanism. Fairness focuses on whether the sequence of transactions before being processed by miners are consistent with the final ordering. The property is used to prevent adversarial manipulation against the actual ordering of transactions in the mempool.

Our proposed attacks do not affect the properties of either consistency or liveness, while only breaking the fairness property. The reason lies in the assumption that we maximally limit the activities of ``adversaries''. Traditional attacks generally empower adversaries with many out-competing abilities, such as cloning an instance~\cite{vincent20poa} or falsifying identities. Even much, the attack may require further requirements on total synchronous network assumptions. In contrast, we try to obey assumptions as they have been defined in the initial stage. We merely allow the adversary (a profitable sealer) to conduct activities that have not been clarified, rather than violating any explicit rules. Since miners' activities have not been seriously monitored or discussed, our attacks can be successfully implemented in many long-term running PoA-engined projects, and instances are listed in Tab.\ref{tab-impact}.

\subsection{Comparisons with Parallel Attacks}

We consider three types of attacks threatening the security of PoA protocols: \textit{51\% attack}, \textit{cloning attack} and \textit{balance attack}. 51\% attack means an adversary can take control over 51\% of sealers. Due to the committee being static and transparent, compromising half of committee members is theoretically feasible when it is on a small scale. Cloning attack \cite{vincent20poa} allows an adversary to clone the instance of a sealer (key pair, metadata). Two equivalent groups containing the same number of sealers are isolated to create hard forks. Similarly, balance attack \cite{natoli2016balance} aims to profitably keep (at least) two branches growing at the same pace by partitioning the network into multiple zones. An adversary can dynamically wander across those balanced zones and select the most profitable one to maximize her revenues. As shown in the attached table, all these attacking strategies rely on a similar attacking strategy: creating double-spending transactions through a hard fork. This will break both the consistency and liveness, destroying the operation from underlying systems. Our attacks differ from these strategies, where a system can still work as normal. 
\begin{wraptable}{r}{5cm}
\begin{threeparttable}
\centering
 \resizebox{\linewidth}{!}{\begin{tabular}{lccc}
    \textit{\textbf{Attack}} & \rotatebox{0}{\textit{\textbf{Consistency}}} & \rotatebox{0}{\textit{\textbf{Liveness}}} & \rotatebox{0}{\textit{\thead{Fairness}}} \\     
    \midrule
    \textit{Type-I}  &  \CheckmarkBold & \CheckmarkBold & \XSolidBrush  \\  
    \textit{Type-II}  & \CheckmarkBold   & \CheckmarkBold & \XSolidBrush\\ 
    51\% &  \XSolidBrush   & \XSolidBrush & N/A  \\ 
    Cloning &  \XSolidBrush  & \XSolidBrush  & N/A \\ 
    Balance &  \XSolidBrush   & \XSolidBrush & N/A  \\ 
    \midrule
  \end{tabular}}
\end{threeparttable}
\end{wraptable}
\noindent It only impacts a single targeted victim, rather than all users participating in the system. Further, the victims may not even be aware of whether they were attacked, namely, frountrun by others or not.

\subsection{Attacking Impacts}
We have explored 2,500+ \textit{in the wild} projects from Github repositories and selected a small set of mainstream projects (10+, see Tab.\ref{tab-impact}) as instances. The reason to select these projects is that most of the current PoA implementations follow very similar protocol designs, or even directly fork the code from other projects. We only narrow down our focus to mainstream implementations or high market cap projects that may be influential to communities. Specifically, we dive into the source code of exemplified projects, detecting vulnerabilities and positioning associated locations. We mainly analyze the \textit{Type-II} attack rather than \textit{Type-I} as the \textit{Type-I} attack can be applied to almost every PoA implementation (which has authority nodes). Here, we conclude our insights.

\noindent\hangindent 1em  \textit{\textbf{Insight 1.}} The \textit{Type-II} attack is effective in most \textit{Clique}-based (including variants) PoA implementations. As discussed before, the current version of \textit{Clique} (in Go-Ethereum) depends on a loose verification mechanism where the edge-turn sealers' identities have not been seriously checked. Merely relying on \textit{difficulty} is not reliable, as a profitable sealer can easily keep sending the $0$-time delays while peer nodes cannot distinguish whether she is a normal edge-turn sealer who legally wins the game or a profitable adversary using falsified methods. 

\noindent\hangindent 1em  \textit{\textbf{Insight 2.}} Many existing projects, even for well-known projects (such as Ethereum Classic \cite{etc-clique}, ConsenSys \cite{ConsenSys-clique}, Binance smart chain \cite{binance-clique}) with high market caps, inherit the code from Go-Ethereum client (Geth). The vulnerabilities detected in these projects are witnessed in the same line of source code (e.g., see Tab.\ref{tab-impact} line No.658). Attacks taking effect in Go-Ethereum will also impact these projects with similar damage. 

\noindent\hangindent 1em \textit{\textbf{Insight 3.}} Several \textit{Clique}-based variants can avoid being frontrun by profitable sealers (\textit{Type-II}) due to the constrictions on sealers' behaviours. In most mainstream consensus algorithms, the behaviours of miners are free of supervised (\textit{a.k.a.} dark forest \cite{ethdkfrst}) where they can reorganize transactions according to their preferences. A clear set of predefined rules for these miners is necessary for avoiding the potential ambiguous activities. These rules can define the methods of sorting transactions, such as relying on paid tips, or transaction appearance.


\noindent\hangindent 1em \textit{\textbf{Insight 4.}} Combining PoA with existing consensus protocols cannot mitigate the ordering attacks. Additional \textit{top-up} algorithms can only resolve the problem on how to enter the committee, rather than how to limit members' activities. For instance, BSC integrates DPoS and PoA~\cite{BSC} to a new hybrid algorithm called Proof of Staked Authority, where validators (equiv. sealers in the context of this work) take turns to produce blocks in a PoA manner (similar to Go-Ethereum’s \textit{Clique}). However, the combined DPoS scheme is merely utilized to select a set of eligible authorities based on their stakes. The person who holds more deposits will earn the opportunity of entering the committee.

\begin{table}[htb!] 
  \centering 
  \caption{Attacking Impacts (towards \textit{Type-II} Attack) }  
  \label{tab-impact}
   \begin{threeparttable}
    \resizebox{\linewidth}{!}{  
  \begin{tabular}[t]{llll}
    \toprule
      \textit{\textbf{Project}} & \textit{\textbf{Client}} & \textit{\textbf{Location}} & \textit{\thead{Notes}} \\ \midrule
      
     \rowcolor{tabyellow} Go-Ethereum  &  Clique  &  \textcolor{blue}{No.293} in \cite{go-clique} & No checks on identity \\  
     
     \rowcolor{tabyellow} BSC & Clique & \textcolor{blue}{No.463} in \cite{binance-clique}  & No checks on identity \\

     \rowcolor{tabyellow} Polygon (MATIC) & Clique & \textcolor{blue}{No.463} in \cite{matic-clique} &   Forked from Go-Ethereum \\ 
     
     \rowcolor{tabyellow} Openethereum & Clique & \textcolor{blue}{No.658} in \cite{openethereum-clique}  &  No checks on identity \\  
     
     \rowcolor{tabyellow} PoA network  &  Clique  & \textcolor{blue}{No.658} in \cite{poa2021} &   Forked from Open Ethereum\\ 
     
     \rowcolor{tabyellow} Ethereum Classic & Clique & \textcolor{blue}{No.463} in  \cite{ConsenSys-clique} & Forked from Go-Ethereum  \\  
     
     \rowcolor{tabyellow} ConsenSys & Clique & \textcolor{blue}{No.463} in  \cite{ConsenSys-clique} & Forked from Go-Ethereum    \\  
     
     \rowcolor{tabyellow} GoChain & Clique &  \textcolor{blue}{No.315}  in \cite{gochain-clique} &  Forked from Go-Ethereum    \\
     
     \rowcolor{tabyellow} Daisy  & Clique & \textcolor{blue}{No.305} in \cite{daisy-clique} & No checks on mining  \\
    
     \rowcolor{tabyellow} Olecoin  & Clique & \textcolor{blue}{No.293} in \cite{olecoin-clique} & Forked from Go-Ethereum   \\ 
     
     \rowcolor{tabyellow} EEX   & Clique & - in \cite{exx-clique} & Claim to use original Clique \\ 
      \rowcolor{tabyellow} HPB & Clique &  \textcolor{blue}{No.126} in \cite{hpb-clique} & No checks on mining \\ 
     
     \midrule
     
     AplaProject  & Clique & - in \cite{alpa-clique} & Strong assumption on behaviors \\

     Tomochain & Clique  & \textcolor{blue}{No.746} in \cite{tomochain-clique} &  Check the identity \\  

     Vechain & Aura  & - in \cite{vechain-clique} & Adopt Parity, based on Aura\\
     
     Kovan & Aura & - in \cite{kovan-clique}  & Ethereum Testnet \\ 
    \bottomrule 
  \end{tabular}
  }
  \begin{tablenotes}
       \footnotesize
       \item[1] Data accessed in November, 2021.
       \item[2] Projects in \colorbox{tabyellow}{\color{black}Yellow} are vulnerable to the attack .
       \item[3] Projects without the background can resist the attack.
     \end{tablenotes}
  \end{threeparttable}     
\end{table}

\subsection{Monetary Loss in Marketcap}

As of November 2021, there are $13,890$ crypto projects existing in decentralized markets, reaching up to a total market cap of $2,882$ billion USD, with a 24h volume of $135$ billion USD [CoinMarketCap, Nov 2021]. This breaks the historical records, and the corresponding market cap becomes the highest value ever. The close relation to finance attracts adversaries to keep their eyes on every feasible attack against these cryptocurrencies. A successfully launched attack can benefit these attackers with thousands of returns. To emphasize the influence on markets, we evaluate the potential monetary loss of PoA projects affected by our attacks.
 
We first count several leading examples (cf. Tab.\ref{tab-impact}). The total market value of impacted projects is over $681, 087$ million USD\footnote{We capture several leading projects with associated market caps here: Ethereum (554,476,066,559 USD), Binance (104,900,840,031 USD), Polygon (12,413,978,032 USD), Ethereum Classic (7,874,100,490 USD), GoChain (46,826,436 USD), Daisy (1,199,666,501 USD), PoA network 179,093,222 USD [As of Nov 15, 2021].}, which mainly comes from Ethereum ($81.4\%$) and Binance ($15.4\%$). Meanwhile, Ethereum and Binance smart chain are also the most prevailing smart contract-enabled platforms that support up to more than $2,028$ decentralized applications (DApps)\footnote{Among these DApps, $1,963$ of them are built on top of Ethereum while $65$ are on Binance smart chain. Data captured from \url{https://www.stateofthedapps.com/dapps}.}. The attacks against these DApps can cause hundreds of times of loss than their basic single-chain values. Despite our attacks merely target to profitable miners without breaking normal system operations, the impact can still result in huge monetary loss, especially in financial DApps on top of these platforms, such as DeFi \cite{qin2020attacking}, NFT \cite{wang2021non}, etc.

Here, we dive into two instances. The first case focuses on DeFi. Pancakeswap \cite{pancakeswap} is a decentralized exchange (DEX) built on top of Binance smart chain. Currently, PancakeSwap reaches up to $28,478$ million USD. The attack towards the underlying BSC will also significantly affect PancakeSwap. Users can trade tokens in PancakeSwap as in a centralized exchange, but differently, no third parties participate in the system. As a typical DeFi product, users can use the liquidity pool \cite{gudgeon2020defi} to exchange different tokens. When some users exchange tokens for another type of tokens, other users can earn rewards by staking tokens in the liquidity pool. At the same time, any exchange causes the slippage to move in a reverse direction. Assume that game players send their trades in this liquidity pool. A frontrunning transaction can be processed at a lower price than others. At this time, a profitable sealer may take advantage of trading at a good price. The second case discusses a real scenario of auctions. Suppose that there is an auction event for many valuable art products (based on NFT \cite{wang2021non}), the winner for this bid can obtain thousands of returns in the future with a high probability. A real-world NFT attacking example can be found at~\cite{nftimpact}. A profitable sealer, with aims to win the bid, can launch order manipulation attacks (either by \textit{Type-I} or \textit{Type-II}) to outpace competitors. This results in the \textit{unfairness} of entire participants.

\section{Countermeasures}
\label{sec-recomd}

In this section, we give countermeasures to mitigate the unfairness issue in PoA protocols. Both recommended methods hold the same goal for guaranteeing a \textit{proper turn} of sealers and transactions.

\smallskip
\noindent\textbf{Intuitive Fix}. As analyzed before, we observe that the \textit{Type-I} unfairness is caused by authorities from the block proposers. It is hard to be fixed without introducing external techniques like using zero-knowledge proof to hide the transaction ordering against miners. In contrast, the \textit{Type-II} unfairness comes from the incautious design of code, especially for the \textit{verification} procedure. An intuitive method of avoiding vulnerability is to ensure that the identities of edge-turn sealers are properly matched before proposing blocks when the leader is absent. Equivalently, modifying the function of \texttt{Verfity}\{...\textit{diff=(difficulty)...}\} in Algm.\ref{alg-proposeclique} line 26 (by adding identification checks)  to \texttt{Verify}\{...\textit{diff=(difficulty, \underline{id$_i$})...}\}, as it is in Algm.\ref{alg-proposeaura} line 20, can work well against the issue. This type of fix is the simplest way to mitigate the \textit{Type-II} attack, resolving the root cause from the source code. The fix can be further applied to the projects that follow similar designs or adopt the original Git-code forked from Ethereum \textit{Clique}. Here, we have also submitted our recommend fix to an \textit{in the wild} project, with details referring to Appendix A.

It is worth mentioning that, however, this type of fix is not a perfect solution because it decreases the system availability to some extent. In original settings, when the leader fails, all the edge-turn sealers produce blocks, and the fast one will become the winner. This is the most economical way for the system. Once adding the identification check, the closest edge-turn sealer will produce the block with a significantly higher accepting probability than others, which obviously decreases the competition among those sealers who can produce blocks. Such a situation is the meaning of availability decreasing. Moreover, adding identification checks towards edge-turn sealers brings more computing workload.

\smallskip
\noindent\textbf{Physical Randomness}. Existing PoA implementations set the \textit{difficulty} in a deterministic way (either as $1$ or $2$). This makes an adversary precisely predicate the position of sealers, especially in the \textit{Clique} algorithm. An example is to further bring physical randomness into the system (see Algm.\ref{alg-randm} in Appendix A, source code in \cite{hpbchaingen}). The physical random numbers are periodically sampled from embedded FPGA hardware \cite{hpb-rdmNum}, which provides completely unpredictable randomness compared to pseudo-randomness (generated by a fixed seed). The project resolves the unfairness issue caused by the \textit{Type-II} attack in two key procedures. Firstly, it adds the verification process of the edge-turn sealers' identities (cf. Algm.\ref{alg-randm} line 18). It is consistent with the solution discussed previously. Secondly, it embeds the hardware-based randomness into the verification step. The edge-turn sealers need to involve the sampled randomness when producing the block, and peered sealers who receive the block should check its validity by sampling the same randomness from the same source (Algm.\ref{alg-randm} line 3/5/16/20). This ensures the following events: (i) the edge-turn sealer is correctly elected within her turn by verifying the identity; and (2) the adversary cannot fake enrolled sealers' identities by cloning open-sourced information because physical randomness is unpredictable, which acts as a one-time pad. The method avoids the threat of the \textit{Type-II} attack and guarantees sealers' fairness of their incoming turns. However, this solution cannot mitigate the general \textit{Type-I} attack.

\smallskip
\noindent\textbf{Crypto-based Solution.} Cryptographic primitives can help to mitigate the unfairness problem in multiple aspects. For the \textit{Type-I} attack, we introduce three types of crypto-based methods to minimize the loss of MEV. The first approach \cite{duan2018beat} employs threshold encryption to distribute the power of a single sealer into multiple cooperated sealers. The second method \cite{duan2017secure} is to use commit-reveal schemes to make miners' behaviours accountable. The third way \cite{stathakopoulou2021adding} leverages the hardware-assisted trusted execution environment to ensure the integrity of program execution. These methods add input causality to the ordering process, improving the cost of behaving profit from miners. 


For the \textit{Type-II} attack, a promising way is to realize the confidential sealer rotation by hiding the identities of sealers. Traditional ways of leader election that rely on the static formula (e.g., Algm.\ref{alg-proposeaura} line 4) is a public procedure to all participants. An adversary can easily predict who will become the leader in a certain round and launch a single-point attack like DDoS. Hiding the turn or identities of sealers can reduce the risk of being compromised by external attackers. To achieve confidentiality, we introduce the technique of verifiable random function (VRF) \cite{micali1999verifiable} as an illustrative instance (see details in Appendix C) to demonstrate how to seal the identities of candidates properly. 

\section{Related Work}
\label{sec-relatedwk}

In this section, we focus on the intrinsic property of PoA in terms of its permissionless setting and security property. Then, we review the \textit{fairness} notion in the context of blockchain. 

\smallskip
\noindent\textbf{BFT in Permissionless Settings.} BFT protocols \cite{CastroL99} are the most prevailing consensus family in distributed systems. BFT protocols provide guarantees of strong safety and instant finality, even under a partial synchronous network. BFT protocols are permissioned, splitting network participants into two types, committee members (perform consensus) and clients (synchronize information). The committee is fixed, in most cases, \textit{static} as well. In contrast, classic blockchain systems use the Nakamoto consensus (PoW and variants) \cite{nakamoto2008bitcoin}, a family of permissionless protocols that rely on a probabilistic way for consistency and better liveness. However, this method is inefficient, as every block has to be buried sufficiently deep to avoid being reversed.  Adapting BFT to permissionless systems is an alternative solution \cite{bridgingBFT} to balance consistency and performance. These systems select a group of authorities via trusted ways to run consensus, same as they act in original designs. One major difference lies in their \textit{dynamic} committees, where members can conditionally join and leave following predefined rules. Exemplified protocols include the PoA in Ethereum, dBFT in NEO \cite{wang2020security}, vote-based BFT in Algorand \cite{gilad2017algorand} and variant PBFT in Thunderella \cite{pass2018thunderella}. Meanwhile, many systems adopt a hybrid or two-layer architecture to further leverage advantages of both consensus such as PoS+PBFT in Tendermint \cite{amoussou2018correctness}, etc.

\smallskip
\noindent\textbf{Security Analysis in PoA.} Angelis et al. \cite{de2018pbft} provide the first discussion of PoA by comparing it with PBFT under the CAP theorem. They point out the threat of inconsistency under an asynchronous network. Shi et al. \cite{shi2019analysis} further explore the situation when an adversary is able to maintain two branches in equal lengths. The malicious sealer only needs to falsify the timestamp when claiming its priority. Based on a similar idea, Parinya et al. \cite{vincent20poa} propose the cloning attack by allowing an adversary sealer to clone her key pair into peered nodes. Committee members isolated in different groups cannot have negotiations, and as a result, hard forks happen without chances to converge. The attack demonstrates that PoA is not secure under common assumptions of partial synchronous network and the presence of malicious sealers. Meanwhile, several analyses utilizing different methodologies also put focus on PoA protocols.  Liu et al. \cite{liu2019mdp} leverage Markov Decision Process to quantitatively measure the balance-off between security and scalability. The proposed framework is applied to VeChainThor's PoA algorithm. Cyril et al. \cite{samuel2021choice} propose a load-balancer middleware to test Ethereum clients, covering Geth, Parity , and Besu (Hyperledger). They evaluate clients from the perspectives of behaviours and performance, and discuss bottlenecks with reasons. Toyoda et al. \cite{toyoda2020function} provide a function-level analysis by developing a Golang’s profiling tool \textit{pprof}. They find the targeted functions that slow the confirmation and hinder the concurrency. Besides, other generic tools for the testing can also be found in \cite{dinh2017blockbench,rouhani2017performance}.

\smallskip
\noindent\textbf{Fairness in Blockchain.} The term of \textit{fairness} refers to block-level security that may threaten the chain stability. It is used in block mining in PoW protocols, indicating that the mining rewards should be proportional to the node's contribution of computing power. A miner cannot perform selfish mining \cite{eyal2014majority} for more payments than its deserved rewards (fare share \cite{kelkar2020order}). In this sense, a set of protocols, including FruitChain \cite{pass2017fruitchains}, SmartPool \cite{luu2017smartpool}, Miller's work \cite{miller2015nonoutsourceable}, have been proposed to achieve \textit{fairness} by strengthening protocols from structure designs to incentive mechanisms. Later, the \textit{fairness} notion focus on transaction-level honesty. Helix et al. \cite{asayag2018fair} propose the way of using threshold encryption to select a random set of pending transactions during the block creation. Meanwhile, several DAG-based blockchain protocols \cite{wang2020sok}, such as Hashgraph \cite{baird2016swirlds} and Prism \cite{bagaria2019prism}, also provide their solutions to sort transactions in a linear order, rather than letting them staying in disordered status \cite{popov2018tangle,wang2021weak}. Further, Kelkar et al. formally define the transaction-ordering fairness in both permissioned \cite{kelkar2020order} and permissionless settings \cite{kelkar2021order}. They present the strict models and related analyses, with aims to prevent adversarial manipulation of the actual transaction ordering. 

\section{Conclusion}
\label{sec-conclu}
PoA algorithms have been adopted by numerous blockchain projects. Their design principles used by different variants are basically the same. However, the lack of restrictions on node behaviours and further authentication of identities (regarding implementation level) make them vulnerable to being tampered within the order of transactions and blocks. These projects reach a market cap of approximately $681,087$ million USD. Corresponding high values will, inevitably, result in severe loss of real money once attacked by adversaries. In this paper, we explore such vulnerabilities, denoted as the \textit{unfairness} issue, existing in PoA protocols. We have classified two types of attacks based on a comprehensive review of 10+ source codes of influential projects among 2,500+ \textit{in the wild} ones. The first attack (transaction-unfairness) breaks the legal order (equiv. first-in-first-out) of transactions, while the second attack (block-unfairness) targets to break the normal rotation of sealers, which can produce high-priority blocks. Both of them merely rely on honest-but-\textit{profitable} sealer assumption, which may happen in the real world with a high probability. We apply our attacks to two most prevailing implementations, including \textit{Aura} and \textit{Clique}, and evaluate their effectiveness. In addition, we propose several ways of remedies by diving into their source code and analysing root causes. To our knowledge, this paper is the first work to investigate the \textit{unfairness} issue of PoA algorithms, enlightening the following designs and providing potential insights to communities.

\medskip
\noindent\textbf{Acknowledgement.}
With the highest importance, we want to express our sincere thanks to expert researchers who first formally propose the notion of \textit{fairness} in the context of blockchain and developers who have publicly released such various PoA implementations. Also, we thank Xinrui Zhang (SUSTech) for her help. Lastly, Qi Wang and Rujia Li are partially supported by the Shenzhen Fundamental Research Programs under Grant No.20200925154814002.

\bibliographystyle{unsrt}
\bibliography{bib.bib}

\begin{thebibliography}{10}

\bibitem{poachains}
Proof-of-authority chains.
\newblock {\em \url{https://openethereum.github.io/Proof-of-Authority-Chains}},
  2016.

\bibitem{de2018pbft}
Stefano De~Angelis, Leonardo Aniello, Roberto Baldoni, Federico Lombardi,
  Andrea Margheri, and Vladimiro Sassone.
\newblock Pbft vs proof-of-authority: Applying the cap theorem to permissioned
  blockchain.
\newblock 2018.

\bibitem{wood2014ethereum}
Gavin Wood et~al.
\newblock Ethereum: A secure decentralised generalised transaction ledger.
\newblock {\em Ethereum Project Yellow Paper}, 151(2014):1--32, 2014.

\bibitem{parity}
Ethereum parity.
\newblock {\em \url{https://www.parity.io/}}, 2021.

\bibitem{geth}
Ethereum geth.
\newblock {\em \url{https://geth.ethereum.org/}}, 2021.

\bibitem{kovan}
Kovan testnet.
\newblock {\em \url{https://kovan-testnet.github.io/website/}}, 2021.

\bibitem{weber2019platform}
Ingo Weber, Qinghua Lu, An~Binh Tran, et~al.
\newblock A platform architecture for multi-tenant blockchain-based systems.
\newblock In {\em 2019 IEEE International Conference on Software Architecture
  (ICSA)}, pages 101--110. IEEE, 2019.

\bibitem{vechain}
Vechain blockchain.
\newblock {\em \url{https://www.vechain.org/}}, 2021.

\bibitem{azure}
Deploy ethereum proof-of-authority consortium solution template on azure.
\newblock {\em
  \url{https://docs.microsoft.com/en-us/azure/blockchain/templates/ethereum-poa-deployment}},
  2021.

\bibitem{barinov2019proof}
Igor Barinov, Vadim Arasev, Andreas Fackler, Vladimir Komendantskiy, Andrew
  Gross, Alexander Kolotov, and Daria Isakova.
\newblock Proof of stake decentralized autonomous organization.
\newblock 2019.

\bibitem{clique}
Clique.
\newblock {\em \url{https://github.com/ethereum/EIPs/issues/225}}, 2021.

\bibitem{BSC}
Binance smart chain.
\newblock {\em
  \url{https://github.com/binance-chain/whitepaper/blob/master/WHITEPAPER.md}},
  2021.

\bibitem{aws-Kaleido}
Launch enterprise-ready blockchain networks on aws in minutes with kaleido—a
  consensys solution.
\newblock {\em
  \url{https://aws.amazon.com/blogs/apn/launch-enterprise-ready-blockchain-networks-on-aws-in-minutes-with-kaleido-a-consensys-solution/}},
  2021.

\bibitem{poa}
Poa.
\newblock {\em \url{https://www.poa.network/}}, 2021.

\bibitem{rinkeby}
Rinkeby: Ethereum testnet.
\newblock {\em \url{https://www.rinkeby.io//#stats}}, 2021.

\bibitem{quorum}
Quorum github repository.
\newblock {\em \url{https://github.com/ConsenSys/quorum}}, 2021.

\bibitem{goerli2022net}
G\.{o}erli network.
\newblock {\em \url{https://goerli.net/}}, 2022.

\bibitem{Apla}
Apla blockchain.
\newblock {\em
  \url{https://apla.readthedocs.io/en/latest/concepts/consensus.html}}, 2021.

\bibitem{hpb-poa}
High performance blockchain (hpb) -- consensus.
\newblock {\em
  \url{https://github.com/hpb-project/go-hpb/tree/master/consensus}}, 2021.

\bibitem{sompolinsky2015secure}
Yonatan Sompolinsky and Aviv Zohar.
\newblock Secure high-rate transaction processing in bitcoin.
\newblock In {\em International Conference on Financial Cryptography and Data
  Security (FC)}, pages 507--527. Springer, 2015.

\bibitem{sompolinsky2013accelerating}
Yonatan Sompolinsky and Aviv Zohar.
\newblock Accelerating bitcoin's transaction processing, fast money grows on
  trees, not chains.
\newblock {\em Cryptology ePrint Archive}, 2013.

\bibitem{kelkar2020order}
Mahimna Kelkar, Fan Zhang, Steven Goldfeder, and Ari Juels.
\newblock Order-fairness for byzantine consensus.
\newblock In {\em Annual International Cryptology Conference (CRYPTO)}, pages
  451--480. Springer, 2020.

\bibitem{kelkar2021order}
Mahimna Kelkar, Soubhik Deb, and Sreeram Kannan.
\newblock Order-fair consensus in the permissionless setting.
\newblock {\em IACR Cryptol. ePrint Arch.}, 2021:139, 2021.

\bibitem{daian2020flash}
Philip Daian, Steven Goldfeder, Tyler Kell, Yunqi Li, Xueyuan Zhao, Iddo
  Bentov, Lorenz Breidenbach, and Ari Juels.
\newblock Flash boys 2.0: Frontrunning in decentralized exchanges, miner
  extractable value, and consensus instability.
\newblock In {\em 2020 IEEE Symposium on Security and Privacy (SP)}, pages
  910--927. IEEE, 2020.

\bibitem{zhou2021high}
Liyi Zhou, Kaihua Qin, Christof~Ferreira Torres, Duc~V Le, and Arthur Gervais.
\newblock High-frequency trading on decentralized on-chain exchanges.
\newblock In {\em 2021 IEEE Symposium on Security and Privacy (SP)}, pages
  428--445. IEEE, 2021.

\bibitem{ethdkfrst}
Robinson Dan and Konstantopoulos Georgios.
\newblock Ethereum is a dark forest.
\newblock 2021.

\bibitem{go-clique}
Clique in go-ethereum.
\newblock {\em
  \url{https://github.com/ethereum/go-ethereum/blob/master/consensus/clique/clique.go}},
  2021.

\bibitem{hpbpr22}
HPB.
\newblock Pr report (hpb).
\newblock {\em \url{https://github.com/hpb-project/go-hpb/pull/82/}}, 2022.

\bibitem{etc-clique}
Consensus in ethereum classic (etc).
\newblock {\em
  \url{https://github.com/etclabscore/core-geth/blob/master/consensus/clique/clique.go}},
  2021.

\bibitem{olecoin-clique}
Consensus in olecoin.
\newblock {\em
  \url{https://github.com/Olecoin/olechain/blob/master/consensus/clique/clique.go}},
  2021.

\bibitem{vincent20poa}
Parinya Ekparinya, Vincent Gramoli, and Guillaume Jourjon.
\newblock The attack of the clones against proof-of-authority.
\newblock In {\em 27th Annual Network and Distributed System Security Symposium
  (NDSS)}, 2020.

\bibitem{wang2020sok}
Qin Wang, Jiangshan Yu, Shiping Chen, and Yang Xiang.
\newblock Sok: Diving into dag-based blockchain systems.
\newblock {\em arXiv preprint arXiv:2012.06128}, 2020.

\bibitem{torres2021frontrunner}
Christof~Ferreira Torres, Ramiro Camino, et~al.
\newblock Frontrunner jones and the raiders of the dark forest: An empirical
  study of frontrunning on the ethereum blockchain.
\newblock In {\em 30th USENIX Security Symposium (USENIX Security)}, pages
  1343--1359, 2021.

\bibitem{natoli2016balance}
Christopher Natoli and Vincent Gramoli.
\newblock The balance attack against proof-of-work blockchains: The r3 testbed
  as an example.
\newblock {\em arXiv preprint arXiv:1612.09426}, 2016.

\bibitem{ConsenSys-clique}
Clique in consensys.
\newblock {\em
  \url{https://github.com/ConsenSys/quorum/blob/master/consensus/clique/clique.go}},
  2021.

\bibitem{binance-clique}
Posa in binance smart chain.
\newblock {\em
  \url{https://github.com/binance-chain/whitepaper/blob/master/WHITEPAPER.md#proof-of-staked-authority}},
  2021.

\bibitem{matic-clique}
Poa in polygon(matic) chain.
\newblock {\em
  \url{https://github.com/maticnetwork/bor/tree/master/consensus/clique}},
  2021.

\bibitem{openethereum-clique}
Clique in openethereum.
\newblock {\em
  \url{https://github.com/openethereum/openethereum/blob/main/crates/ethcore/src/engines/clique/mod.rs}},
  2021.

\bibitem{poa2021}
Poa network.
\newblock {\em
  \url{https://github.com/poanetwork/openethereum/blob/main/ethcore/src/engines/clique/mod.rs}},
  2021.

\bibitem{gochain-clique}
Clique in gochain.
\newblock {\em
  \url{https://github.com/gochain-io/go-ethereum/blob/master/consensus/clique/clique.go}},
  2021.

\bibitem{daisy-clique}
Consensus in daisy.
\newblock {\em
  \url{https://github.com/ivoras/daisy/blob/18f27b8469b8684f19dff59dc39c5457e45e4fd2/blockchain.go#L206}},
  2021.

\bibitem{exx-clique}
Proposals in exx.
\newblock {\em \url{https://cex.ethereum-express.com/files/whitepaper.pdf}},
  2021.

\bibitem{hpb-clique}
Clique in hpb.
\newblock {\em
  \url{https://github.com/hpb-project/go-hpb/blob/master/consensus/prometheus/chain_verification.go}},
  2021.

\bibitem{alpa-clique}
Confirmation procedures in aplaproject.
\newblock {\em
  \url{https://github.com/AplaProject/go-apla/blob/8e17a21423289ea4aeb0d9021cfb46e283acf217/packages/block/block.go}},
  2021.

\bibitem{tomochain-clique}
Consensus in tomochain.
\newblock {\em
  \url{https://github.com/tomochain/tomochain/blob/722d85a8318ba4fcfaffd91d9855acb29217ecaa/consensus/posv/posv.go#L746}},
  2021.

\bibitem{vechain-clique}
Validation procedures in vechain.
\newblock {\em
  \url{https://github.com/vechain/thor/blob/master/consensus/validator.go}},
  2021.

\bibitem{kovan-clique}
Proposals in kovan.
\newblock {\em \url{https://github.com/kovan-testnet/proposal}}, 2021.

\bibitem{qin2020attacking}
Kaihua Qin, Liyi Zhou, Benjamin Livshits, and Arthur Gervais.
\newblock Attacking the defi ecosystem with flash loans for fun and profit.
\newblock In {\em International Conference on Financial Cryptography and Data
  Security (FC)}, pages 3--32. Springer, 2021.

\bibitem{wang2021non}
Qin Wang, Rujia Li, Qi~Wang, and Shiping Chen.
\newblock Non-fungible token (nft): Overview, evaluation, opportunities and
  challenges.
\newblock {\em arXiv preprint arXiv:2105.07447}, 2021.

\bibitem{pancakeswap}
Pancakeswap.
\newblock {\em \url{https://pancakeswap.finance/}}, 2021.

\bibitem{gudgeon2020defi}
Lewis Gudgeon, Sam Werner, Daniel Perez, and William~J Knottenbelt.
\newblock Defi protocols for loanable funds: Interest rates, liquidity and
  market efficiency.
\newblock In {\em Proceedings of the 2nd ACM Conference on Advances in
  Financial Technologies (AFT)}, pages 92--112, 2020.

\bibitem{nftimpact}
Copeland Tim.
\newblock Mistake sees \$69,000 cryptopunk sold for less than a cent.
\newblock {\em
  \url{https://www.theblockcrypto.com/post/113546/mistake-sees-69000-cryptopunk-sold-for-less-than-a-cent}},
  2022.

\bibitem{hpbchaingen}
Chain generation process in hpb.
\newblock {\em
  \url{https://github.com/hpb-project/go-hpb/blob/master/consensus/prometheus/chain_generation.go}},
  2021.

\bibitem{hpb-rdmNum}
High performance blockchain (hpb) -- hardware generated random number.
\newblock {\em
  \url{https://github.com/hpb-project/hardware-random-number-case}}, 2021.

\bibitem{duan2018beat}
Sisi Duan, Michael~K Reiter, and Haibin Zhang.
\newblock Beat: Asynchronous bft made practical.
\newblock In {\em Proceedings of the 2018 ACM SIGSAC Conference on Computer and
  Communications Security (AsiaCCS)}, pages 2028--2041, 2018.

\bibitem{duan2017secure}
Sisi Duan, Michael~K Reiter, and Haibin Zhang.
\newblock Secure causal atomic broadcast, revisited.
\newblock In {\em 2017 47th Annual IEEE/IFIP International Conference on
  Dependable Systems and Networks (DSN)}, pages 61--72. IEEE, 2017.

\bibitem{stathakopoulou2021adding}
Chrysoula Stathakopoulou, Signe R{\"u}sch, Marcus Brandenburger, and Marko
  Vukoli{\'c}.
\newblock Adding fairness to order: Preventing front-running attacks in bft
  protocols using tees.
\newblock In {\em 2021 40th International Symposium on Reliable Distributed
  Systems (SRDS)}, pages 34--45. IEEE, 2021.

\bibitem{micali1999verifiable}
Silvio Micali, Michael Rabin, and Salil Vadhan.
\newblock Verifiable random functions.
\newblock In {\em 40th Annual Symposium on Foundations of Computer Science
  (FOCS)}, pages 120--130. IEEE, 1999.

\bibitem{CastroL99}
Castro Miguel and Liskov Barbara.
\newblock Practical byzantine fault tolerance.
\newblock In {\em Proceedings of the Third USENIX Symposium on Operating
  Systems Design and Implementation (OSDI)}, pages 173--186, 1999.

\bibitem{nakamoto2008bitcoin}
Satoshi Nakamoto.
\newblock Bitcoin: A peer-to-peer electronic cash system.
\newblock {\em Decentralized Business Review}, page 21260, 2008.

\bibitem{bridgingBFT}
Viswanath Pramod.
\newblock Lecture 16: Bridging bft protocols with the longest chain protocol.
\newblock {\em
  \url{https://courses.grainger.illinois.edu/ece598pv/sp2021/lectureslides2021/ECE_598_PV_course_notes16_v2.pdf}},
  2021.

\bibitem{wang2020security}
Qin Wang, Jiangshan Yu, Zhiniang Peng, Van~Cuong Bui, Shiping Chen, Yong Ding,
  and Yang Xiang.
\newblock Security analysis on dbft protocol of neo.
\newblock In {\em International Conference on Financial Cryptography and Data
  Security (FC)}, pages 20--31. Springer, 2020.

\bibitem{gilad2017algorand}
Yossi Gilad, Rotem Hemo, Silvio Micali, Georgios Vlachos, and Nickolai
  Zeldovich.
\newblock Algorand: Scaling byzantine agreements for cryptocurrencies.
\newblock In {\em Proceedings of the 26th Symposium on Operating Systems
  Principles (SOSP)}, pages 51--68. ACM, 2017.

\bibitem{pass2018thunderella}
Rafael Pass and Elaine Shi.
\newblock Thunderella: Blockchains with optimistic instant confirmation.
\newblock In {\em Annual International Conference on the Theory and
  Applications of Cryptographic Techniques (EUROCRYPT)}, pages 3--33. Springer,
  2018.

\bibitem{amoussou2018correctness}
Yackolley Amoussou-Guenou, Antonella Del~Pozzo, Maria Potop-Butucaru, and Sara
  Tucci-Piergiovanni.
\newblock Correctness and fairness of tendermint-core blockchains.
\newblock {\em arXiv preprint arXiv:1805.08429}, 2018.

\bibitem{shi2019analysis}
Elaine Shi.
\newblock Analysis of deterministic longest-chain protocols.
\newblock In {\em 2019 IEEE 32nd Computer Security Foundations Symposium
  (CSF)}, pages 122--12213. IEEE, 2019.

\bibitem{liu2019mdp}
Xuefeng Liu, Gansen Zhao, Xinming Wang, Yixing Lin, Ziheng Zhou, Hua Tang, and
  Bingchuan Chen.
\newblock Mdp-based quantitative analysis framework for proof of authority.
\newblock In {\em 2019 International Conference on Cyber-Enabled Distributed
  Computing and Knowledge Discovery (CyberC)}, pages 227--236. IEEE, 2019.

\bibitem{samuel2021choice}
Cyril~Naves Samuel, Severine Glock, Fran{\c{c}}ois Verdier, and Patricia
  Guitton-Ouhamou.
\newblock Choice of ethereum clients for private blockchain: Assessment from
  proof of authority perspective.
\newblock In {\em 2021 IEEE International Conference on Blockchain and
  Cryptocurrency (ICBC)}, pages 1--5. IEEE, 2021.

\bibitem{toyoda2020function}
Kentaroh Toyoda, Koji Machi, Yutaka Ohtake, and Allan~N Zhang.
\newblock Function-level bottleneck analysis of private proof-of-authority
  ethereum blockchain.
\newblock {\em IEEE Access}, 8:141611--141621, 2020.

\bibitem{dinh2017blockbench}
Tien Tuan~Anh Dinh, Ji~Wang, Gang Chen, Rui Liu, Beng~Chin Ooi, and Kian-Lee
  Tan.
\newblock Blockbench: A framework for analyzing private blockchains.
\newblock In {\em Proceedings of the 2017 ACM International Conference on
  Management of Data (SIGMOD)}, pages 1085--1100, 2017.

\bibitem{rouhani2017performance}
Sara Rouhani and Ralph Deters.
\newblock Performance analysis of ethereum transactions in private blockchain.
\newblock In {\em 2017 8th IEEE International Conference on Software
  Engineering and Service Science (ICSESS)}, pages 70--74. IEEE, 2017.

\bibitem{eyal2014majority}
Ittay Eyal and Emin~G{\"u}n Sirer.
\newblock Majority is not enough: Bitcoin mining is vulnerable.
\newblock In {\em International conference on Financial Cryptography and Data
  Security (FC)}, pages 436--454. Springer, 2014.

\bibitem{pass2017fruitchains}
Rafael Pass and Elaine Shi.
\newblock Fruitchains: A fair blockchain.
\newblock In {\em Proceedings of the ACM Symposium on Principles of Distributed
  Computing (PODC)}, pages 315--324, 2017.

\bibitem{luu2017smartpool}
Loi Luu, Yaron Velner, Jason Teutsch, and Prateek Saxena.
\newblock Smartpool: Practical decentralized pooled mining.
\newblock In {\em 26th USENIX Security Symposium (USENIX Security)}, pages
  1409--1426, 2017.

\bibitem{miller2015nonoutsourceable}
Andrew Miller, Ahmed Kosba, Jonathan Katz, and Elaine Shi.
\newblock Nonoutsourceable scratch-off puzzles to discourage bitcoin mining
  coalitions.
\newblock In {\em Proceedings of the 22nd ACM SIGSAC Conference on Computer and
  Communications Security (CCS)}, pages 680--691, 2015.

\bibitem{asayag2018fair}
Avi Asayag, Gad Cohen, Ido Grayevsky, et~al.
\newblock A fair consensus protocol for transaction ordering.
\newblock In {\em 2018 IEEE 26th International Conference on Network Protocols
  (ICNP)}, pages 55--65. IEEE, 2018.

\bibitem{baird2016swirlds}
Leemon Baird.
\newblock The swirlds hashgraph consensus algorithm: Fair, fast, byzantine
  fault tolerance.
\newblock {\em Swirlds Tech Reports SWIRLDS-TR-2016-01, Tech. Rep}, 2016.

\bibitem{bagaria2019prism}
Vivek Bagaria et~al.
\newblock Prism: Deconstructing the blockchain to approach physical limits.
\newblock In {\em ACM SIGSAC Conference on Computer and Communications Security
  (CCS)}, pages 585--602, 2019.

\bibitem{popov2018tangle}
Serguei Popov.
\newblock The tangle.
\newblock {\em White paper}, 1(3), 2018.

\bibitem{wang2021weak}
Qin Wang and Rujia Li.
\newblock A weak consensus algorithm and its application to high-performance
  blockchain.
\newblock In {\em IEEE INFOCOM 2021-IEEE Conference on Computer
  Communications}, pages 1--10. IEEE, 2021.

\bibitem{AlgorandSOSP2017}
Yossi Gilad, Rotem Hemo, Silvio Micali, Georgios Vlachos, and Nickolai
  Zeldovich.
\newblock Algorand: Scaling byzantine agreements for cryptocurrencies.
\newblock In {\em Proceedings of the 26th Symposium on Operating Systems
  Principles (SOSP)}, pages 51--68, 2017.

\end{thebibliography}

\section*{Appendix A. HPB Algorithms}
We provide the core procedures (mainly focus on the \texttt{Propose} and \texttt{Verify} steps) of HPB \cite{hpb-poa}, an PoA-engined project (see Algm.\ref{alg-randm}). Reasons to capture these two functions are: (i) Both \texttt{Propose} and \texttt{Verify} functions invoke the hardware-based randomness from the same source (Algm.\ref{alg-randm} line 3/16). This ensures that the identified items (e.g., \textit{trace}/\textit{tag}) can keep consistent, proving that the incoming sealer correctly enters her legal turn. (ii) The root cause of \textit{Type-II} vulnerability is the absence of identity checks towards incoming sealers. We thereby suggest adding associated check procedures (e.g., Algm.\ref{alg-randm} line 16-21, 27). Similar to the solution of \textbf{Intuitive Fix}, the key revision is to add the verification of sealers' identities into the current version. The difference lies in the usage of physical random numbers sampled from FPGA hardware \cite{hpb-rdmNum}. The hardware-based randomness contributes to the system from two sides. Firstly, it is embedded into the system as a checking parameter, guaranteeing that every peer node (sample \textit{tag}) can verify the block producer's identity (sample \textit{trace}). Secondly, randomness is used in the sealer rotation step (Algm.\ref{alg-randm} line 5), where any participant cannot predict who will become the incoming block producer. This fix has been submitted to the HPB official team. Note that after discussing with the team, they have finally accepted part of our suggestions -- fixing the synchronization problem, to ensure the smooth operations of the \textit{difficulty} verification (cf.~\cite{hpbpr22}).

\begin{algorithm}
\caption{HPB with Hardware Randomness}
\label{alg-randm}
\begin{algorithmic}[1]
\Procedure{Propose}{$sealer_i$} \Comment{\textcolor{blue}{propose a block}} 
\While{$(\mathsf{true})$} 
\State $trace \gets sample(Hardware) $
\State \Comment{\textcolor{blue}{sample randomness from hardware}} 
\State $\textbf{if} \, (trace+1) \mod |sealers| = i,$\, \textbf{then}
\State \Comment{\textcolor{blue}{miner position}}
\State $\qquad b.weight = 2$ \Comment{\textcolor{blue}{in-turn sealing}} 
\State $ \qquad broadcast(block)$  \Comment{\textcolor{blue}{send block immediately}} 
\State $\textbf{else}$
\State $\qquad b.weight = 1$ \Comment{\textcolor{blue}{edge-turn sealing}} 
\State $ \qquad sleep(.) $  \Comment{\textcolor{blue}{send block with delay}}
\State $ \qquad  broadcast(block)$ 
\EndWhile\label{euclidendwhile}

\State
\Procedure{Verify}{$sealer_i$}
\Comment{\textcolor{blue}{verify a block}}
\State $\textit{tag} \gets sample(Hardware)$
\State  \Comment{\textcolor{blue}{sample tag from the same hardware}} 
\State \textit{diff} $=$ \textit{calcDifficulty(header.number, sealer)}
\State \Comment{\textcolor{blue}{compute difficulty}}
\State \textit{miner} $=$ \textit{calcMiner(header.number, tag)}
\State \Comment{\textcolor{blue}{compute miner in this round}}
\State $\textbf{if} \, (trace == tag),$\, \textbf{then}
\State $ \quad \textbf{if} \, (\textit{diff}\, == nil),$ \textbf{then} \Comment{\textcolor{blue}{not nil}}
\State $\qquad \textbf{return}$  $\mathsf{false}$ 
\State $ \quad \textbf{else if} \, (\textit{diff} \,\, !=  1 \, \&\& \,\textit{diff} \,\,!= 2)$ \textbf{then} \Comment{\textcolor{blue}{1 or 2}}
\State $\qquad \textbf{return}$ $\mathsf{false}$
\State $ \quad \textbf{else if} \, (\textit{diff} \,\, ==  1 \, \&\& \,\textit{sealer} \,\,!= miner)$ \textbf{then} \Comment{\textcolor{red}{patch}}
\State $\qquad \textbf{return}$ $\mathsf{false}$
\State $ \quad \textbf{else}$ \textbf{then}
\State $\qquad \textbf{return}$ $\mathsf{true}$ \Comment{\textcolor{blue}{block is valid}}
\EndProcedure

\EndProcedure
\end{algorithmic}
\end{algorithm}

\section*{Appendix B. Notations}
In this part, we highlight several terms used in this paper. We use very easy-understand notations and try to keep consistent with previous similar definitions in concurrent studies such as \cite{vincent20poa,de2018pbft}. Meanwhile, we further provide more precise definitions than existing statements in terms of sealers' behaviours (\textit{honest} and \textit{benign}). Our work is exactly based on these slight differences.

\begin{table}[htb!]
  \centering 
  \caption{Notations, consistent with \cite{vincent20poa}}  
  \label{tab-notation}
    \resizebox{\linewidth}{!}{  
  \begin{tabular}[t]{ccc p{6.2cm}}
    \toprule
      \thead{Symbol} & \thead{Role} &   \thead{Alias}  &\thead{Functionalities} \\ \midrule
     $N_i$  &  Sealer  & Out-of-turn sealer & Committee members, cannot produce blocks \\ \midrule 
     $N_L$ & Leader & In-turn sealer & Produce the block in the highest priority \\  \midrule
     $N_e$ & Sealer & Edge-turn sealer & Produce the block in the secondary priority \\  \midrule
     $f(\centerdot)$ & \multicolumn{2}{c}{Formula} & Rules deciding how to select the leader. \\  \midrule
     $B_i$ &  \multicolumn{2}{c}{Out-of-turn block} & Blocks created by out-of-turn sealers. \\ \midrule
     $B_L$ & \multicolumn{2}{c}{In-of-turn block} & 
     Blocks produced by in/edge-turn sealers.    \\ 
     \midrule
    - & \multicolumn{2}{c}{Honest sealer} & The sealer who obey predefined rules, but can act arbitrary for out-of-scope rules, such as being profitable.  \\ 
    \midrule
     - & \multicolumn{2}{c}{Benign sealer} & The sealer who act honestly without own willingness, faithfully following the predefined rules.\\
    \bottomrule 
  \end{tabular}
  }
\end{table}

\section*{Appendix C. Instance Using VRF}

\textit{Confidential Election via VRF}. Verifiable random function (VRF)~\cite{micali1999verifiable} is a typical public-key pseudo-random function used to prove that the outputs with proofs matches its associated inputs. The owner who holds the secret key can generate the proof from an input string, while others can verify two types of facts by using the public key and the proof: (i) whether the output is created by the owner who holds the secret key; (ii) whether the output is correctly calculated from the associated input. To be noted, Algorand \cite{AlgorandSOSP2017} has also adopted VRF in its cryptographic sortition algorithm to randomly select the sealer while additionally hiding her identity. Here, we present the main steps of constructing VRF-based sealer rotation. Notations are defined as follows: $s$ is a negotiated seed for which corresponds to a specific block height; $pk$ and $sk$ are the sealer's key pair;  $hs$ is the random hash generated by the incoming sealer; $\pi$ is the evidence for the generated hash.

\begin{figure}[!]
    \centering
    \includegraphics[width=\linewidth]{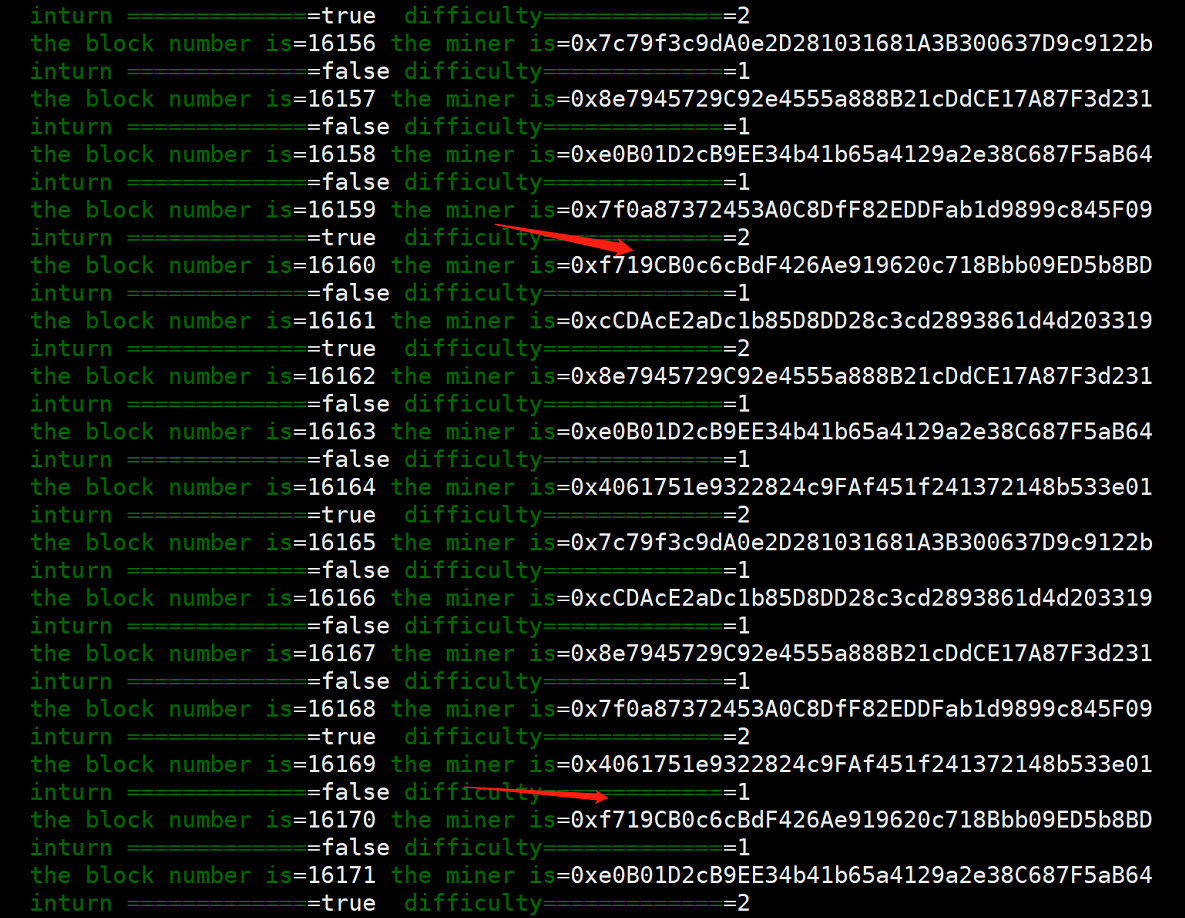}
   \caption{Screenshot: A normal case on running the \textit{Clique} client}
    \label{fig-poa-normal}
\end{figure}

\begin{figure}[!]
    \centering
    \includegraphics[width=\linewidth]{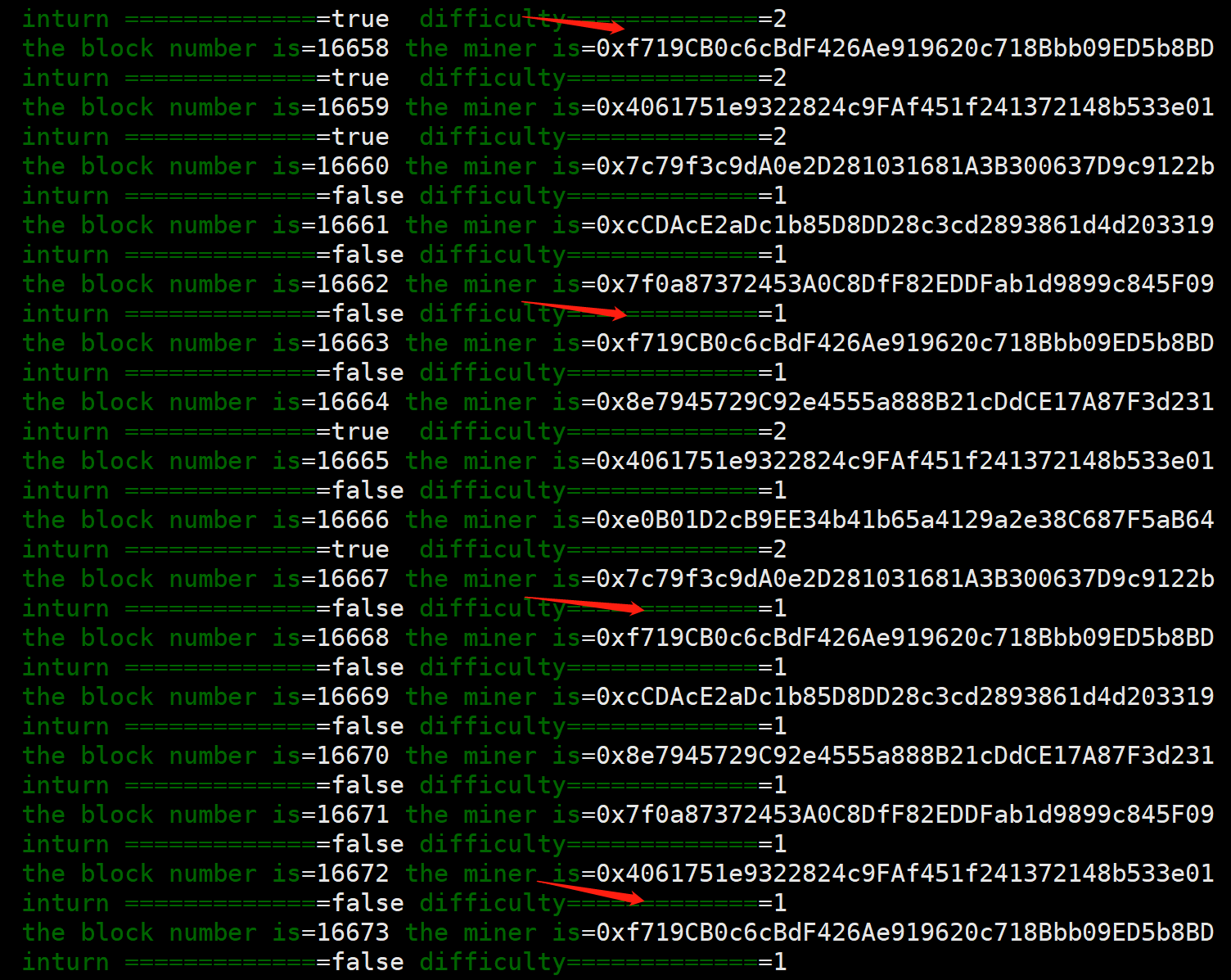}
   \caption{Screenshot:  Launching the \textit{Type-II} attack on \textit{Clique} client}
    \label{fig-poa-attack}
\end{figure}

\begin{itemize}
    \item[-] $\mathsf{VRFKeyGen}(\lambda) \to (sk, pk):$ The algorithm takes as input a security parameter $\lambda$, and outputs the secret/public key pair $(sk, pk)$.
    \item[-] $\mathsf{VRFHash}(sk,s) \to hs:$ The algorithm takes as input $sk$ and a seed $s$, and outputs a fixed-length hash $hs$.
    \item[-] $\mathsf{VRFProve}(sk,s) \to \pi:$ The algorithm takes as input $sk$, $s$, and outputs the proof $\pi$ for $hs$.
    \item[-] $\mathsf{VRFVerify}(pk,s,hs,\pi) \to {0/1}:$ The algorithm takes as input $(pk, m, hs, \pi)$, and outputs the verification result, a boolean variable $0$ or $1$ ($0/1$ for invalid/valid, respectively).
\end{itemize}

Based on the properties of VRF, the proposed confidential scheme obtains additional benefits from three aspects. Firstly, VRF is pseudo-randomness, where an adversary cannot effectively distinguish the output of $\mathsf{VRFHash}$ from a random string without the knowledge of corresponding $pk$ and $\pi$. This enables the sealer to be hidden from the public. Secondly, VRF is unique where one input can only produce a unique output, guaranteeing that attackers cannot falsify the identity of the sealer. Thirdly, VRF is collision-resistant where each output cannot be produced by two different inputs, ensuring adversaries cannot compromise benign sealers. 

\section*{Appendix D. Screenshots of \textit{Type-II} Attack in Our Experiments}


Fig.\ref{fig-poa-normal} and Fig.\ref{fig-poa-attack} show the screenshot of a normal case on running the normal \textit{Clique} clients and a malicious client, respectively. In the normal case, the address $0xf719...$ is assumed to be owned by a benign sealer. In the attacking case, such a benign sealer changes her role to an honest-but-profitable sealer with the same address. In the first experiment, such a sealer operates the program based on a normal Geth client. We observe that only one block has been produced after the in-turn sealer abandons to mine, which means that this sealer cannot frontrun the block. In contrast, in the second experiment, the sealer is supposed to be a profitable sealer and starts to conduct the arbitrage strategy based on a maliciously modified client. We can see that the address $0xf719...$ in Fig.\ref{fig-poa-attack}  (with \textit{difficulty} of 1) appears four times, frontrunning other edge-turn sealers in each rotation round. The results demonstrate that the profit-pursuing sealer has successfully launched the \textit{Type-II} attack. Here, we still keep the leader (with \textit{difficulty} of 2) in the network, but can omit it when counting the result.

\end{document}